\documentclass{aa}     
\usepackage{graphicx}
 
\def\cc{\,{\rm cm^{-3}}}
\def\cm2{\,{\rm cm^{-2}}}
\def\kms{\,{\rm {km\,s^{-1}}}}
\def\kkms{\,{\rm {K\,km\,s^{-1}}}}

\def\h2{\,{\rm H_{2}}}
\def\13co{\,{\rm ^{13}CO}}
\def\co{\,{\rm ^{12}CO}}

\def\aua{{\rm A\&A} }
\def\auas{{\rm A\&AS} }
\def\apj{{\rm ApJ} }
\def\aj{{\rm AJ} }
\def\apjs{{\rm ApJS} }
\def\apjl{{\rm ApJL} }

\def\mnras{{\rm MNRAS} }
\def\pasj{{\rm PASJ} }

\begin{document} 
 
\title{Results of the ESO-SEST Key Programme on CO in the Magellanic Clouds}
 
\subtitle{IX. The giant LMC HII region complex N~11} 
 
\author{F.P. Israel
	\inst{1},
        Th. de Graauw 
        \inst{2},
        L.E.B. Johansson
	\inst{3},
        R.S. Booth
	\inst{3}
        F. Boulanger
	\inst{4,5},
	G. Garay
	\inst{6},
	M.L. Kutner
	\inst{7},
	J. Lequeux
	\inst{8},
	L.-A. Nyman
	\inst{3, 9}
	\and M. Rubio
	\inst{6}
} 

\offprints{F.P. Israel} 
 
\institute{Sterrewacht Leiden, P.O. Box 9513, 2300 RA Leiden, the Netherlands 
\and Laboratorium voor ruimteonderzoek, SRON, Postbus 800, 9700 AV Groningen,
     the Netherlands
\and Onsala Space Observatory, S-439-92 Onsala, Sweden
\and Radioastronomie, Ecole Normale Superieure, 24 rue Lhomond, F-75231, Paris 
     CEDEX 05, France
\and Institut d'Astrophysique Spatiale, Bat. 120, Universit\'e de Paris-XI, 
     F-91045 Orsay CEDEX, France
\and Departamento de Astronomia, Universidad de Chile, Casilla 36-D, 
     Santiago, Chile 
\and no idea, somewhere
\and LERMA, Observatoire de Paris, 61 Av. de l'Observatoire, F-75014 Paris,
     France
\and European Southern Observatory, Casilla 19001, Santiago 19, Chile
}

\date{
Received ????; accepted ????}
 
\abstract{ 
The second-brightest star formation complex in the Large Magellanic Cloud, 
N~11, was surveyed extensively in the $J$=1--0 transition of $\co$. In this
paper we present maps and a catalogue containing the parameters of 29 
individual molecular clouds in the complex, although more may be present.
The distribution of molecular gas in the N~11 complex is highly structured. 
In the southwestern part of N~11, molecular clouds occur in a ring or shell 
surrounding the major OB star association LH~9. In the 
northeastern part, a chain of molecular clouds 
delineates the rim of one of the so-called supergiant shells in the LMC.
There appears to be very little diffuse molecular gas in between the 
individual well-defined clouds, especially in the southwestern ring. Most
of the clouds have dimensions only little larger than those of the
survey beam, i.e diameters of 25 pc or less. A subset of the clouds 
mapped in  $J=1-0 \co$ transition was also observed in the $J=2-1 \co$ 
transition, and in the corresponding transitions of $\13co$. Clouds 
mapped in $J=2-1 \co$ with a twice higher angular resolution show 
further, clear substructure. The elements of this substructure, however, 
have dimensions once again comparable to those of the mapping beam. For 
a few clouds, sufficient information was available to warrant an attempt 
at modelling their physical parameters. They contain fairly warm 
($T_{\rm kin} = 60 - 150$ K) and moderately dense ($n_{\h2} = 3000 \cc$)
gas. The northeastern chain of CO clouds, although lacking in diffuse
intercloud emission, is characteristic of the more quiescent regions
of the LMC, and appears to have been subject to relatively little
photo-processing. The clouds forming part of the southwestern shell
or ring, however, are almost devoid of diffuse intercloud emission, 
and also exhibit other characteristics of an extreme photon-dominated
region (PDR). 
} 

\authorrunning{F.P. Israel et al.}
\titlerunning{CO in LMC-N~11}

\maketitle

\section{Introduction} 

\begin{figure*}[]
\unitlength1cm
\begin{minipage}[]{17.9cm}
\resizebox{17.7cm}{!}{\includegraphics*{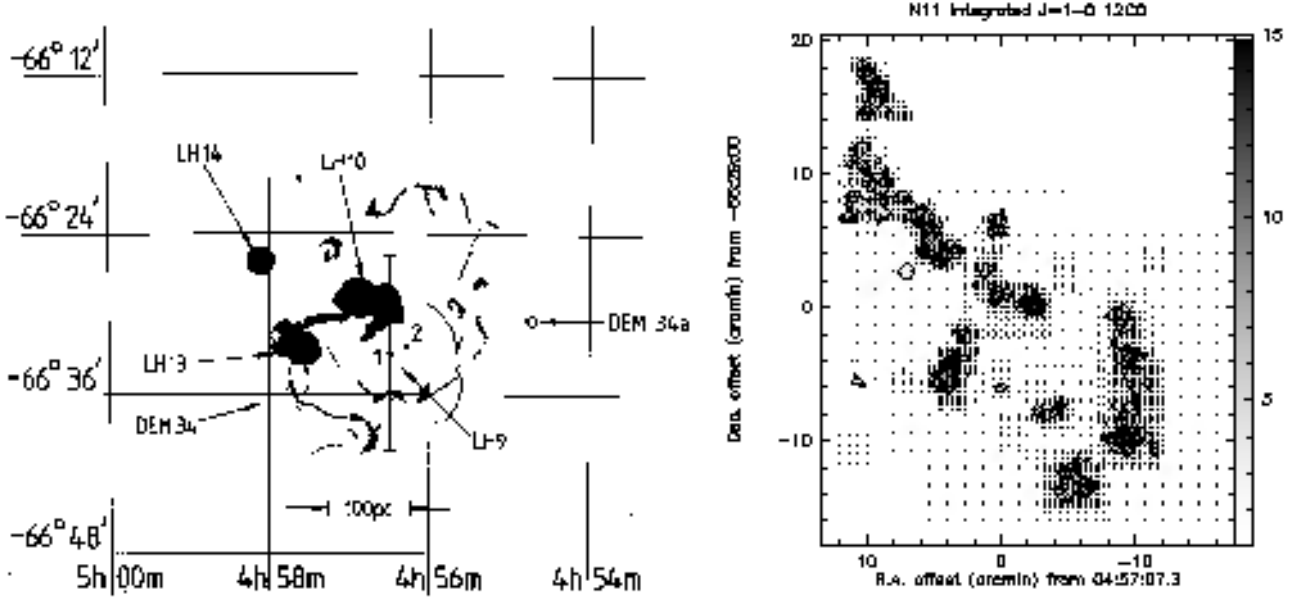}}
\end{minipage}
\caption[]{Left: N~11 sketch from Meaburn et al. (1989). H$\alpha$ emission
is marked by dark zones; OB associations are indicated by their LH
numbers (see text). Right: Overall distribution of integrated $J$=1--0 $\co$ 
emission in N~11 on the same scale.
}
\label{n11overall}
\end{figure*} 

\begin{table*}
\caption[]{Catalogue of CO clouds in N~11}
\begin{flushleft}
\begin{tabular}{rcccrccccccccc}
\hline
\noalign{\smallskip}
No.   & N11$^{a}$ & LI-LMC$^{b}$ & LMC-B$^{c}$ & \multicolumn{2}{c}{Cloud Center$^{d}$} & \multicolumn{4}{c}{Peak $J$=1-0 CO Parameters} & \multicolumn{4}{c}{Peak $J$=2-1 CO Parameters}  \\
      &       &             &       & $\Delta\alpha$ & $\Delta\delta$     & T$_{\rm mb}$ & $\Delta$V & $V_{\rm LSR}$ & $I_{\rm CO}$ & T$_{\rm mb}$ & $\Delta$V & $V_{\rm LSR}$ & $I_{\rm CO}$ \\
      &       &        &       &  $(')$         & $(')$              & (K) &\multicolumn{2}{c}{($\kms$)} & ($\kkms$)   & (K) &\multicolumn{2}{c}{($\kms$)} & ($\kkms$) \\
\noalign{\smallskip}
\hline
\noalign{\smallskip}
  1 &     & 190 &           & -10.0 &  -5.0 & 1.25 & 2.7 & 280.8 &  3.6$\pm$0.5 &       &     &       & \\
  2 &  H  & 190 &           & - 9.5 &  -3.7 & 1.57 & 2.7 & 276.3 &  4.5$\pm$0.6 &       &     &       & \\
  3 &     & 192 &           &  -9.5 &  -8.2 & 1.39 & 2.9 & 284.4 &  4.2$\pm$0.4 &       &     &       & \\
  4 &  I  & 192 &           &  -9.4 & -10.0 & 2.56 & 5.7 & 279.0 & 15.6$\pm$2.0 &       &     &       & \\
  5 & (G) & 195 &           &  -8.5 &  -0.5 & 2.21 & 4.0 & 272.9 &  9.4$\pm$0.9 &       &     &       & \\
  6 &     & 205 &           &  -6.3 & -13.5 & 2.61 & 2.5 & 277.7 &  6.8$\pm$1.0 &       &     &       & \\
  7 &     & 205 &           &  -5.0 & -12.5 & 2.04 & 1.9 & 280.8 &  4.2$\pm$0.6 &  3.27 & 5.0 & 278.4 & 17.5$\pm$0.5 \\
  8 &  F  & 214 & 0456-6636 &  -4.4 &  -7.4 & 2.42 & 2.6 & 276.6 &  6.6$\pm$0.8 &  2.68 & 3.0 & 276.4 &  8.4$\pm$0.4 \\
  9 &  F  & 214 & 0456-6636 &  -3.0 &  -8.0 & 1.46 & 2.6 & 268.6 &  4.0$\pm$0.6 &       &     &       & \\
 10 &  B  & 217 & 0456-6629 &  -2.5 &     0 & 2.47 & 6.1 & 285.6 & 16.0$\pm$1.7 &  5.53 & 5.8 & 285.4 & 34.1$\pm$0.4 \\
 11 &  A  & 226 &           &   0.3 &   1.0 & 1.33 & 4.5 & 277.1 &  6.4$\pm$0.7 &  3.13 & 4.7 & 276.8 & 15.9$\pm$0.5 \\
 12 &  J  & 229 &           &   0.3 &   6.0 & 1.56 & 4.4 & 279.6 &  7.2$\pm$1.0 &       &     &       & \\
 13 &     & 226 &           &   1.0 &   2.2 & 1.39 & 3.1 & 283.1 &  4.6$\pm$0.6 &  1.53 & 2.7 & 278.8 &  4.1$\pm$0.5 \\
 14 &  C  & 243 & 0457-6632 &   2.9 &  -2.3 & 0.96 & 4.0 & 280.2 &  4.4$\pm$0.9 &  3.32 & 4.0 & 279.6 & 14.3$\pm$0.4 \\
 15 &  D  & 248 & 0457-6632 &   3.7 &  -4.3 & 3.00 & 3.8 & 280.7 & 12.0$\pm$2.0 &  3.82 & 4.0 & 280.8 & 16.3$\pm$0.5 \\
 16 &  E  & 251 & 0458-6626 &   4.6 &   3.8 & 1.50 & 5.8 & 268.1 &  9.3$\pm$1.3 &       &     &       & \\
 17 &  E  & 251 & 0458-6626 &   4.2 &   4.2 & 1.46 & 1.7 & 275.2 &  2.6$\pm$0.4 &       &     &       & \\
 18 &  E  &     & 0458-6626 &   5.3 &   5.5 & 2.53 & 3.9 & 271.2 & 10.9$\pm$1.0 &  2.91 & 4.6 & 271.0 & 14.3$\pm$0.4 \\
 19 &     &     &           &   6.4 &   6.8 & 1.54 & 3.7 & 271.3 &  6.0$\pm$0.9 &       &     &       & \\
 20 &     &     &           &   7.4 &   8.3 & 0.57 & 8.3 & 275.4 &  6.0$\pm$0.9 &       &     &       & \\
 21 &     &     &           &   8.7 &   9.2 & 1.14 & 4.8 & 277.0 &  5.8$\pm$0.7 &       &     &       & \\
 22 &     &     &           &   8.7 &   9.7 & 1.32 & 2.0 & 284.1 &  2.8$\pm$0.9 &       &     &       & \\
 23 &     & 268 &           &   9.0 &   7.5 & 0.58 & 8.0 & 274.7 &  4.9$\pm$0.6 &       &     &       & \\
 24 &     & 268 &           &  11.0 &   7.7 & 1.47 & 3.1 & 278.8 &  3.3$\pm$1.0 &       &     &       & \\
 25 &     &     & 0458-6616 &  10.2 &  12.2 & 0.89 & 3.0 & 273.1 &  2.8$\pm$0.6 &       &     &       & \\
 26 &     &     & 0458-6616 &  10.8 &  11.1 & 1.85 & 2.0 & 281.0 &  3.9$\pm$0.6 &       &     &       & \\
 27 &     & 266 &           &   9.8 &  14.7 & 2.01 & 3.9 & 278.4 &  8.5$\pm$0.0 &       &     &       & \\
 28 &     & 271 &           &   9.4 &  16.4 & 2.87 & 3.9 & 276.8 & 11.3$\pm$0.4 &  1.59 & 4.1 & 277.1 &  6.9$\pm$0.4 \\
 29 &     & 271 &           &  10.0 &  17.4 & 2.49 & 3.2 & 275.4 &  8.5$\pm$0.4 &  2.78 & 4.6 & 274.7 & 13.6$\pm$0.5 \\
\noalign{\smallskip}
\hline
\end{tabular}
\end{flushleft}
Notes: 
$^{a}$ Henize (1956) designation 
$^{b}$ IRAS source: Schwering $\&$ Israel (1990). 
$^{c}$ Radio continuum source: Filipovic et al. (1996).
$^{d}$ Offsets refer to central mapping position 
(epoche 1950.0)
$\alpha_{\rm o} = 04^{h}57^{m}07.^{s}3$,
$\delta_{\rm o} = -66^{\rm o}29'00''$.
\end{table*}

\begin{figure*}
\unitlength1cm
\begin{minipage}[b]{18.0cm}
\resizebox{18.0cm}{!}{\rotatebox{0}{\includegraphics*{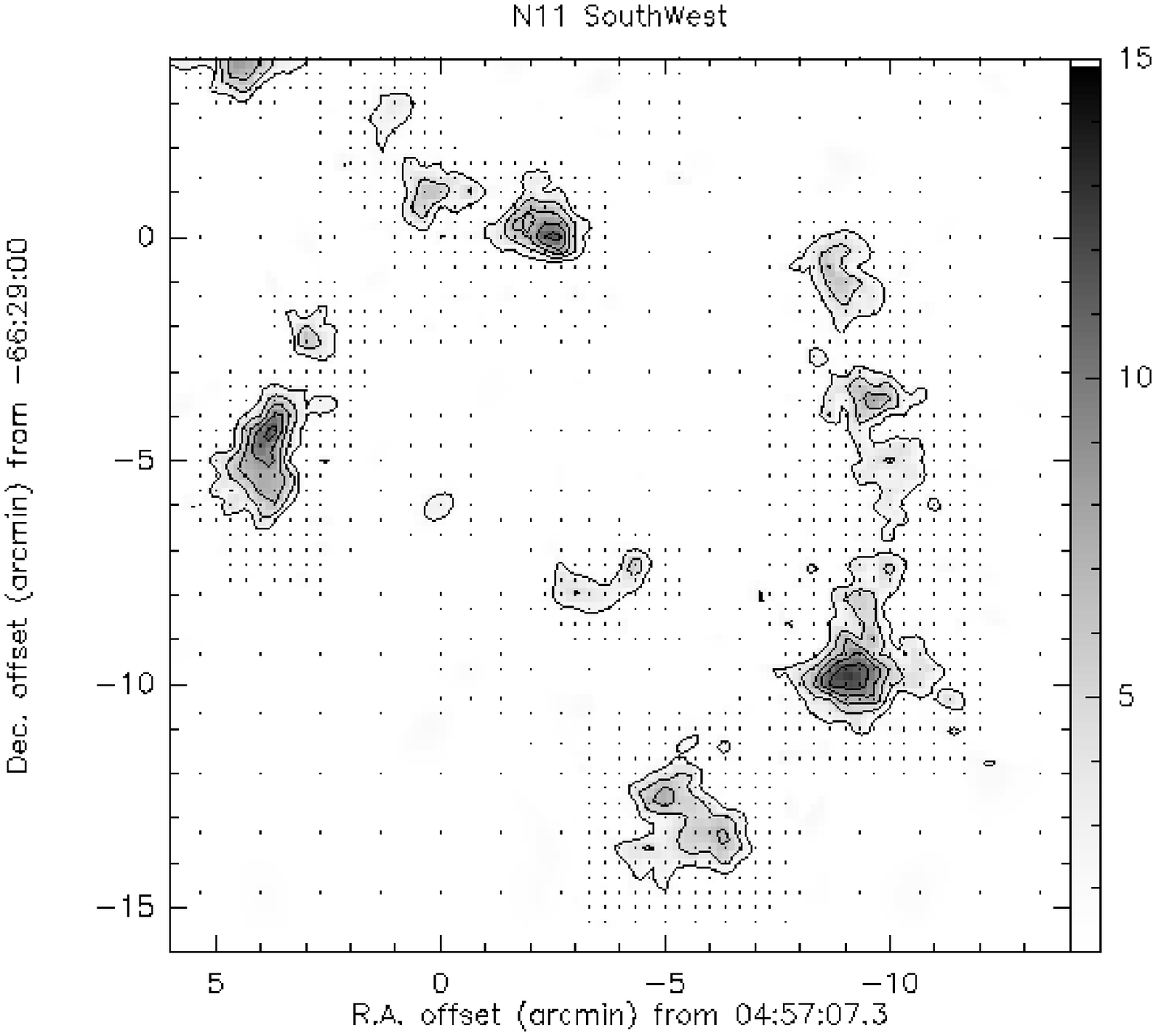}}}
\end{minipage}
\caption[]{Detailed map of the integrated $J$=1--0 $\co$ emission in 
the southwestern part of the N11 complex, showing the `ring' of CO
clouds associated with OB associations LH~9, LH~10 and LH~13.
Positions sampled are indicated by dots. In this Figure, and
in the following, grey scales indicated at the right of the panel are
integrated antenna temperatures. The contours, however, are chosen such 
that both the first contour and the contour interval correspond to 
3 $\kkms$ in main-beam brightness temperature.
}
\label{n11sw}
\end{figure*} 

The ESO SEST Key Programme was established to investigate the
molecular gas in the nearest neighbours to the Milky Way, the
Magellanic Clouds. Considerations pertinent to this programme
were given by Israel et al. (1993; Paper I). Following ESO's
discontinuation of the concept of Key Programmes, the observational 
programme was ended in 1995, although the processing of data obtained
has continued. In this paper we present the results of observations 
of the HII region complex N~11 (Henize 1956), located in the 
northwestern corner of the Large Magellanic Cloud (LMC).  After 
30 Doradus with its retinue of HII regions, supernova remnants and 
dark clouds, this complex is the second-brightest in the LMC.
CO observations of the former, also made within the context of 
the Key Programme, have been published by Johansson et al. (1998) and 
Kutner et al. (1997). 

The N~11 complex is also known as DEM 34 (Davies et al. 1976), and has
an overall diameter of about 45$'$ , corresponding to a linear extent of
705 pc for an assumed LMC distance of 54 kpc (Westerlund, 1990, but see
Walker 1999). In Fig.\ref{n11overall} we present a sketch map of the
optical nebulosity. In the west, N~11 contains the small supernova 
remnant N~11L (= DEM 34a). From the main body of the N~11 complex, a 
loop of HII regions and more diffuse H$\alpha$ emission extends to the 
northeast. This loop delineates the eastern half of LMC supergiant shell 
SGS-1 (Meaburn 1980) which has a diameter of about a kiloparsec and is
centered on OB association LH~15 (Lucke $\&$ Hodge 1970 -- not marked in
Fig.~\ref{n11overall}).

\begin{figure*}
\unitlength1cm
\begin{minipage}[b]{18.0cm}
\resizebox{18.0cm}{!}{\rotatebox{0}{\includegraphics*{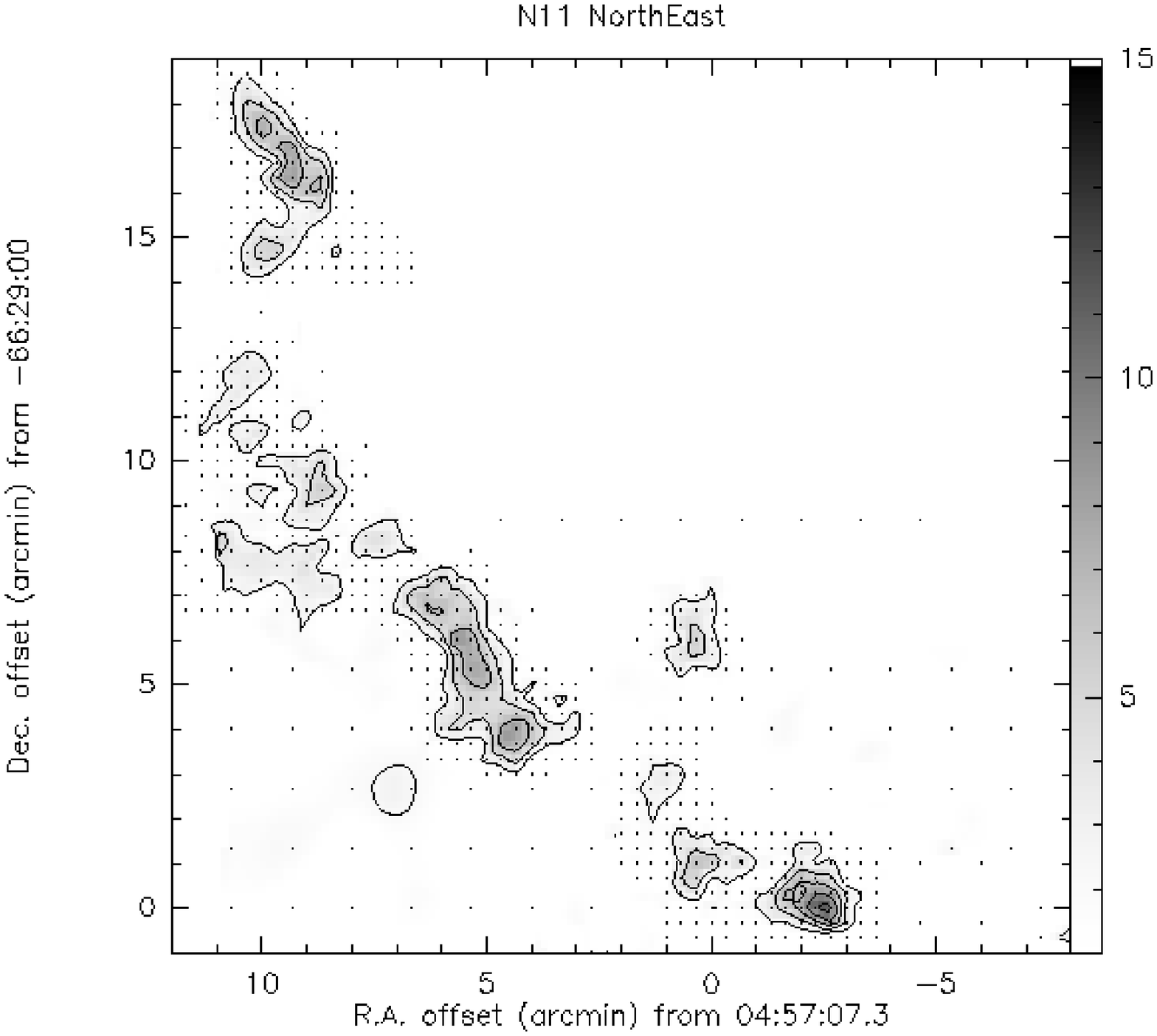}}}
\end{minipage}
\caption[]{Detailed map of the integrated $J$=1--0 $\co$ emission in 
the northeastern part of the N11 complex, showing the `chain' of CO
clouds associated with OB association LH~14 and supergiant shell
SGS~1; otherwise as Fig.~2.
}
\label{n11ne}
\end{figure*} 

\begin{figure*}[]
\unitlength1cm
\begin{minipage}[b]{17.8cm}
\resizebox{17.82cm}{!}{\rotatebox{0}{\includegraphics*{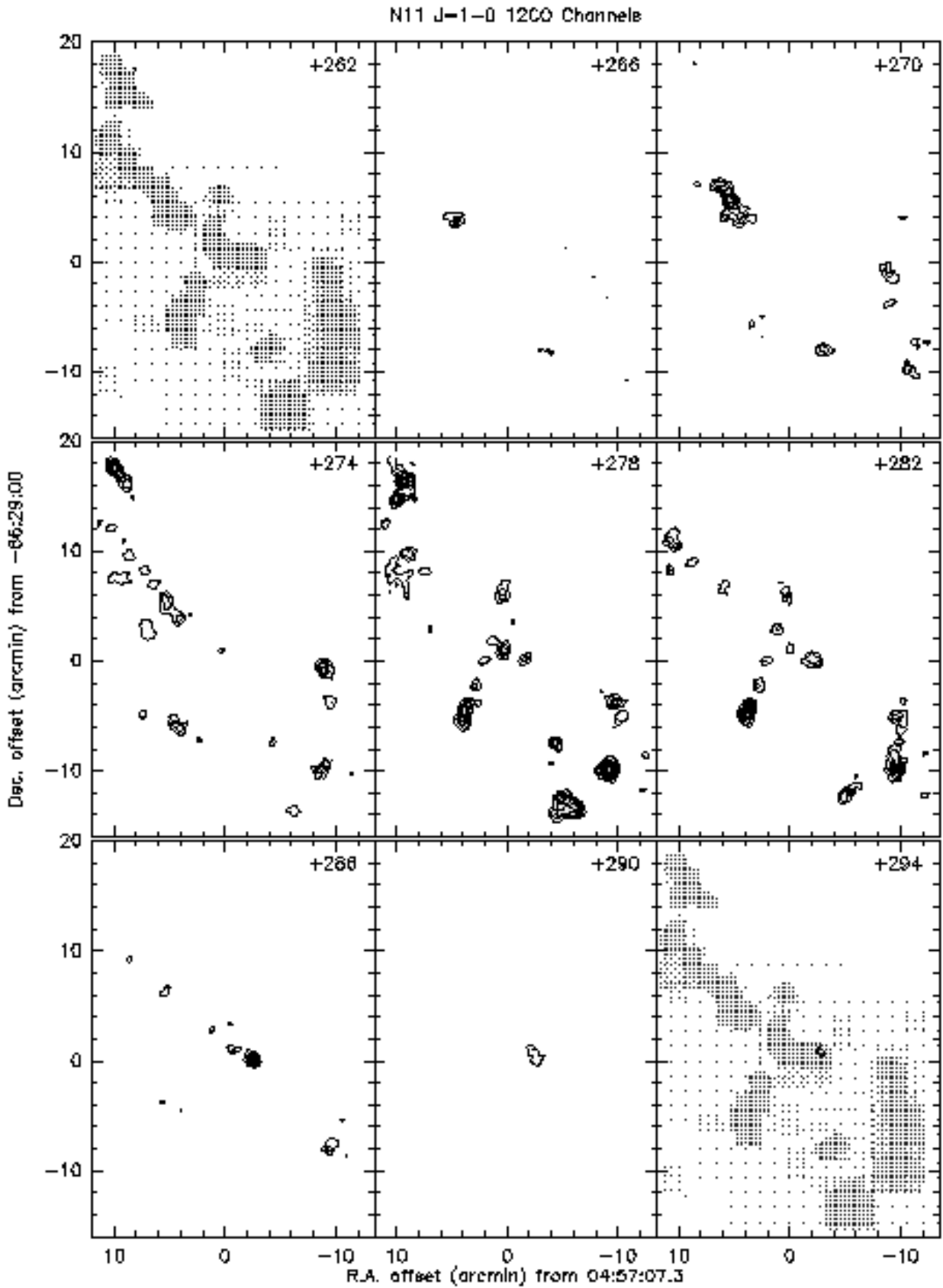}}}
\end{minipage}
\caption{N11 $J$=1-0 $\co$ channel maps. Number in top right corner
indicates central velocity $V_{\rm LSR}$. Positions sampled are indicated 
in the maps centered on +262 and +294 $\kms$. CO emission is integrated
in bins of 4 $\kms$ width. First contour and contour interval correspond to
$\int T_{\rm mb}{\rm d}V = 0.7 \kkms$. 
}
\label{n11chans}
\end{figure*}

\begin{figure*}[]
\unitlength1cm
\begin{minipage}[]{17.5cm}
\begin{minipage}[]{8.75cm}
\resizebox{8.5cm}{!}{\rotatebox{270}{\includegraphics*{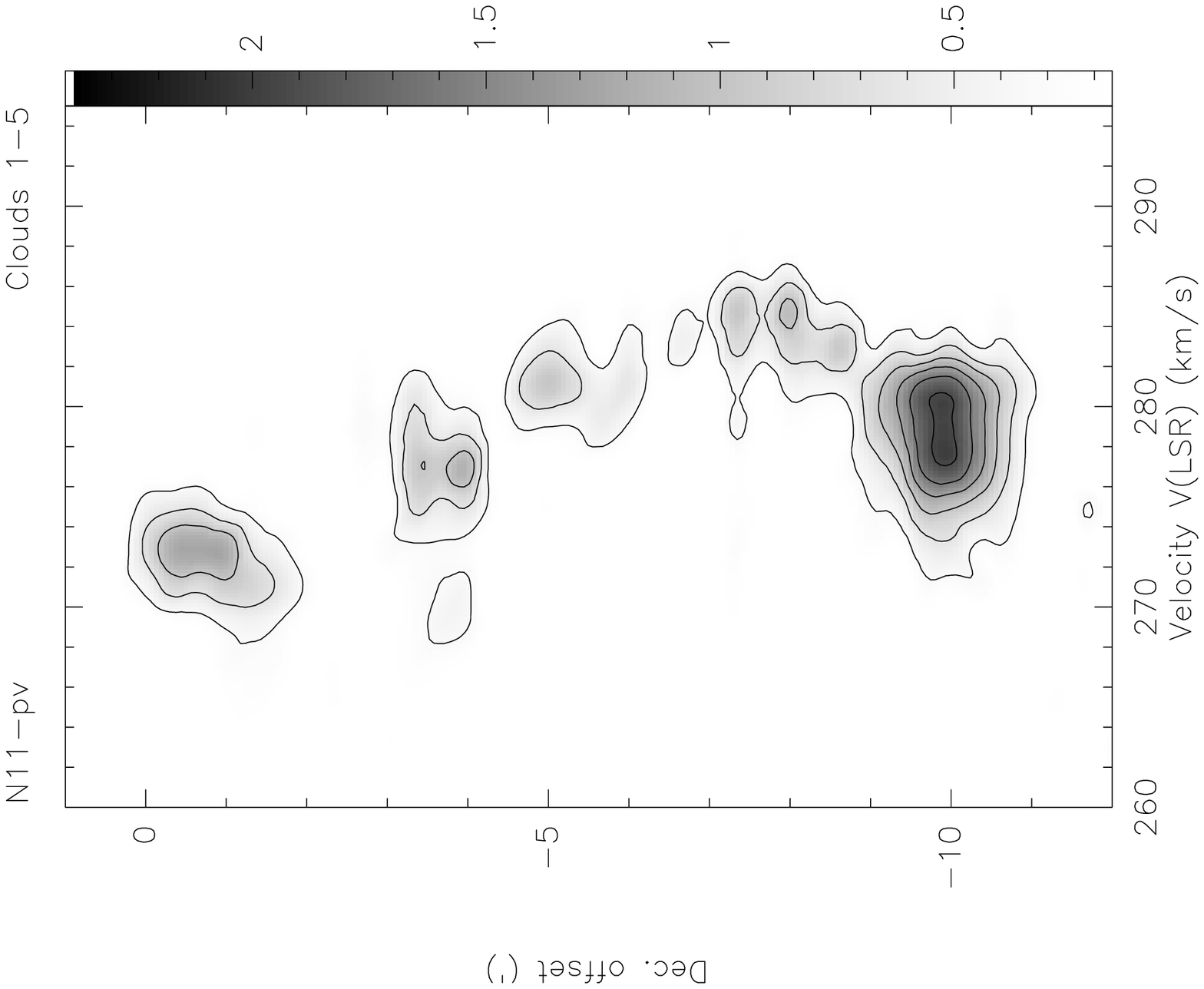}}}
\resizebox{8.5cm}{!}{\rotatebox{270}{\includegraphics*{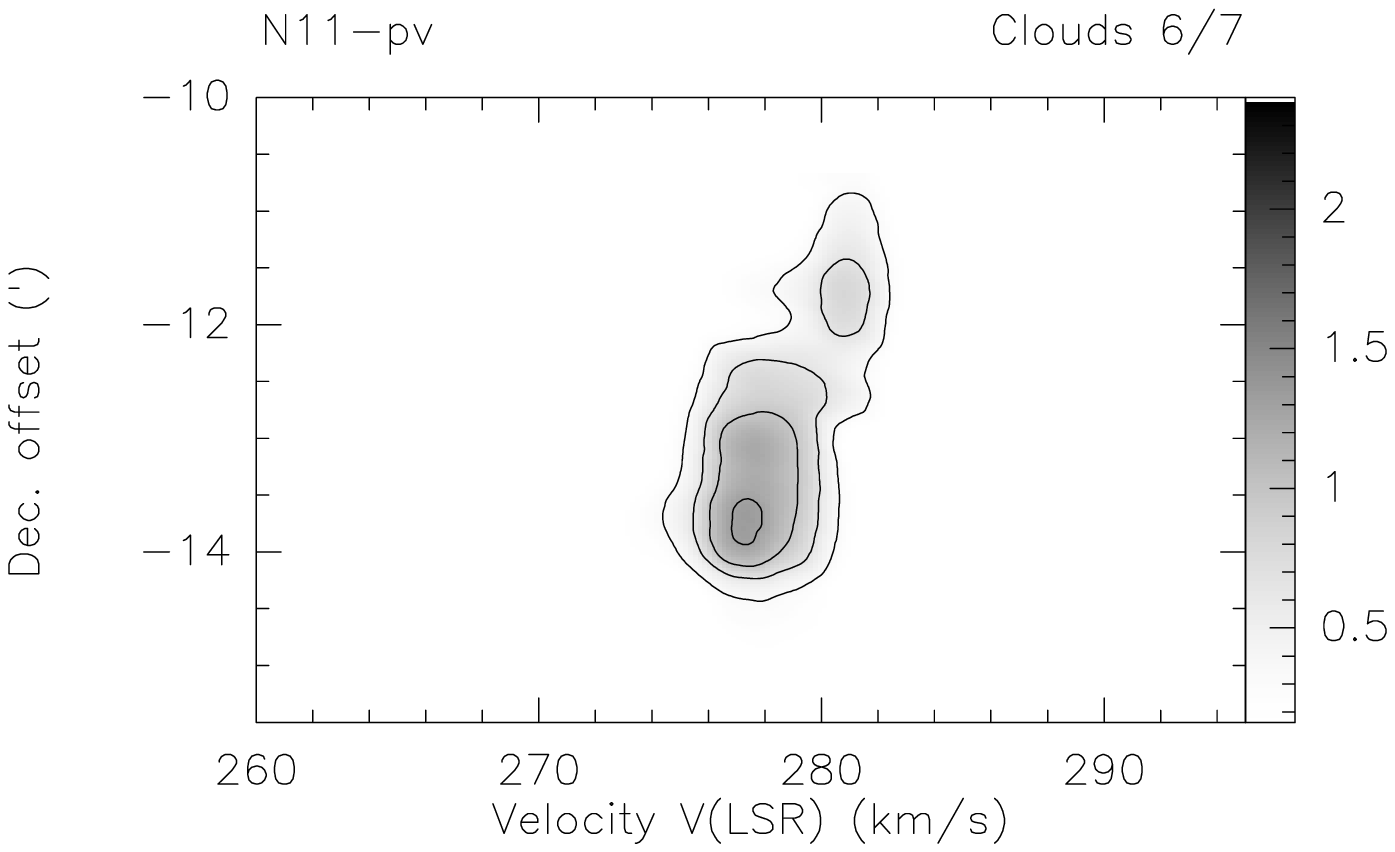}}}
\resizebox{8.5cm}{!}{\rotatebox{270}{\includegraphics*{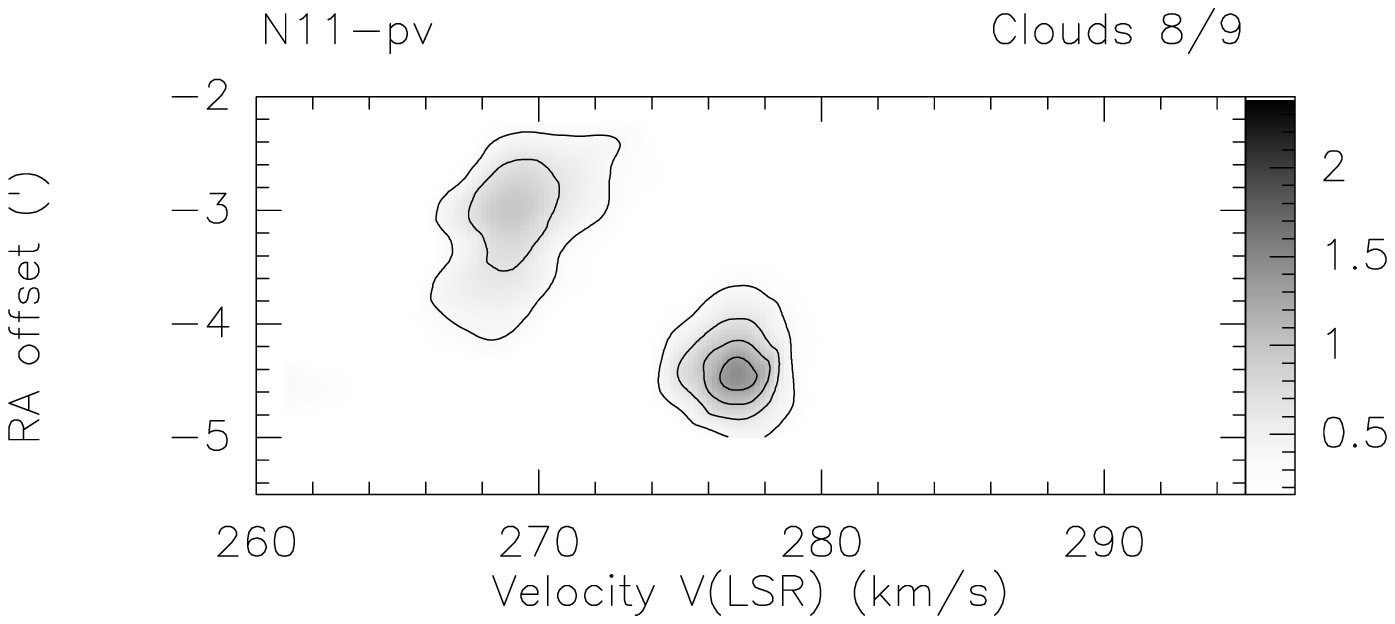}}}
\end{minipage}
\begin{minipage}[]{8.75cm}
\resizebox{8.5cm}{!}{\rotatebox{270}{\includegraphics*{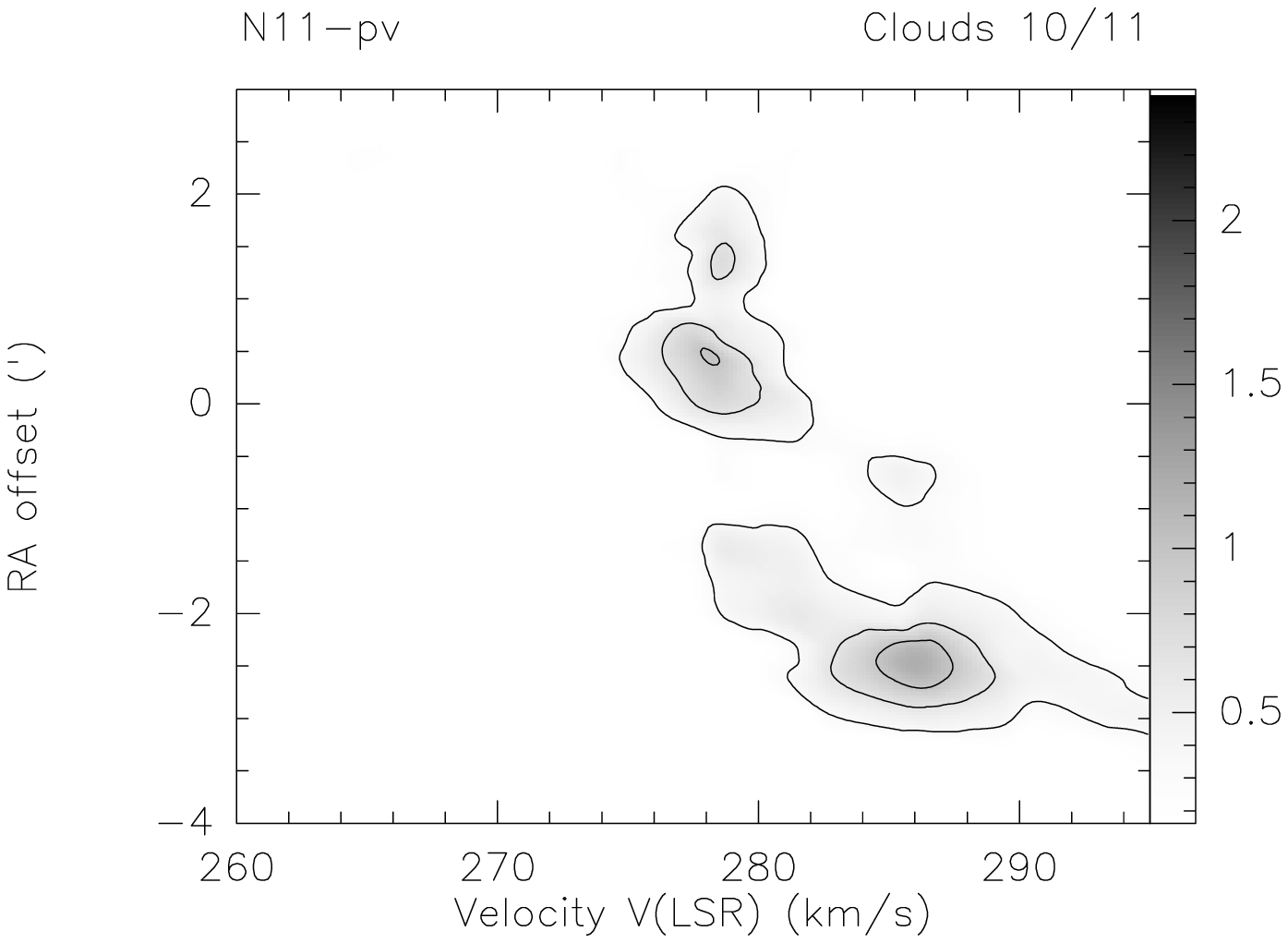}}}
\resizebox{8.5cm}{!}{\rotatebox{270}{\includegraphics*{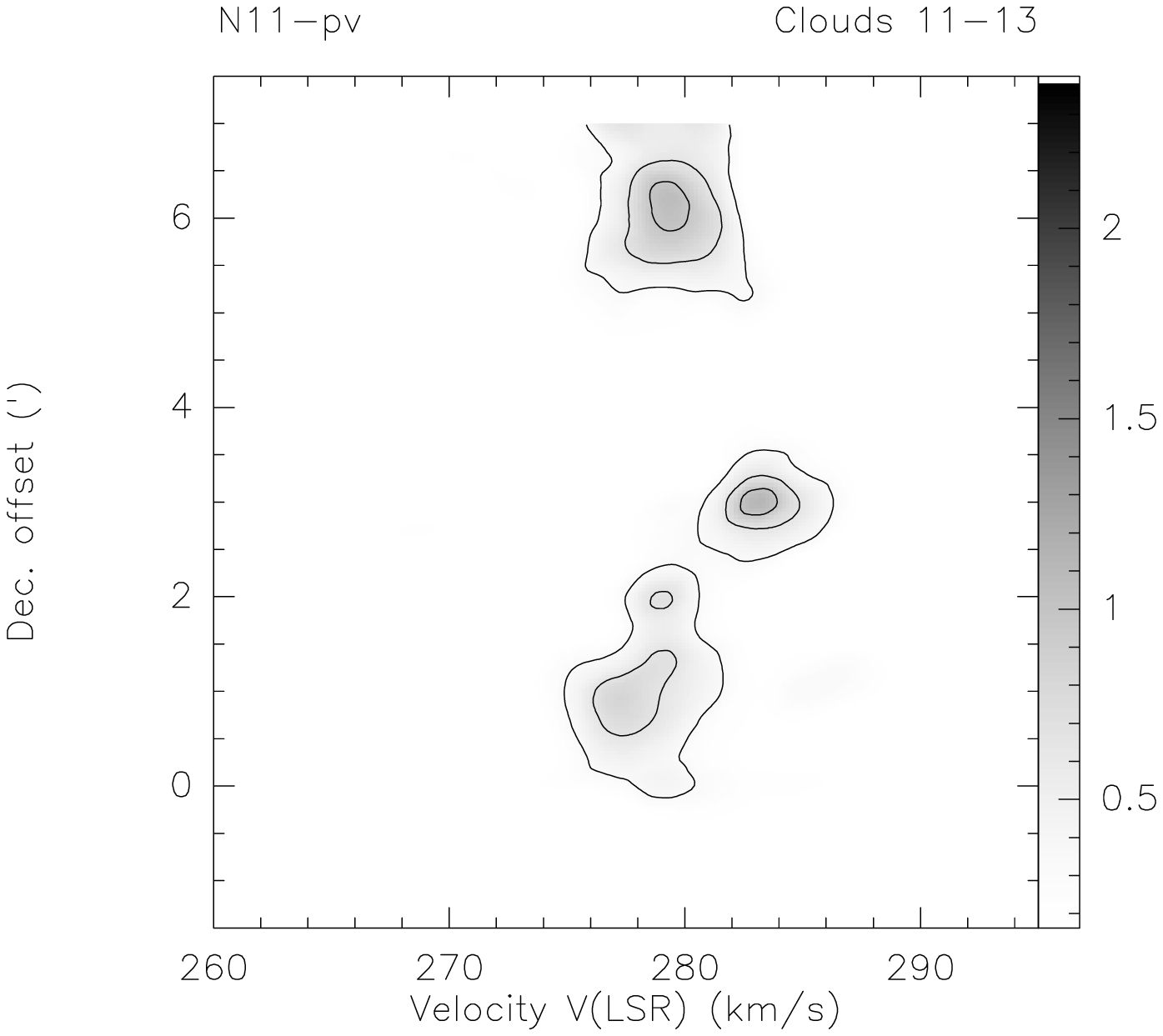}}}
\resizebox{8.5cm}{!}{\rotatebox{270}{\includegraphics*{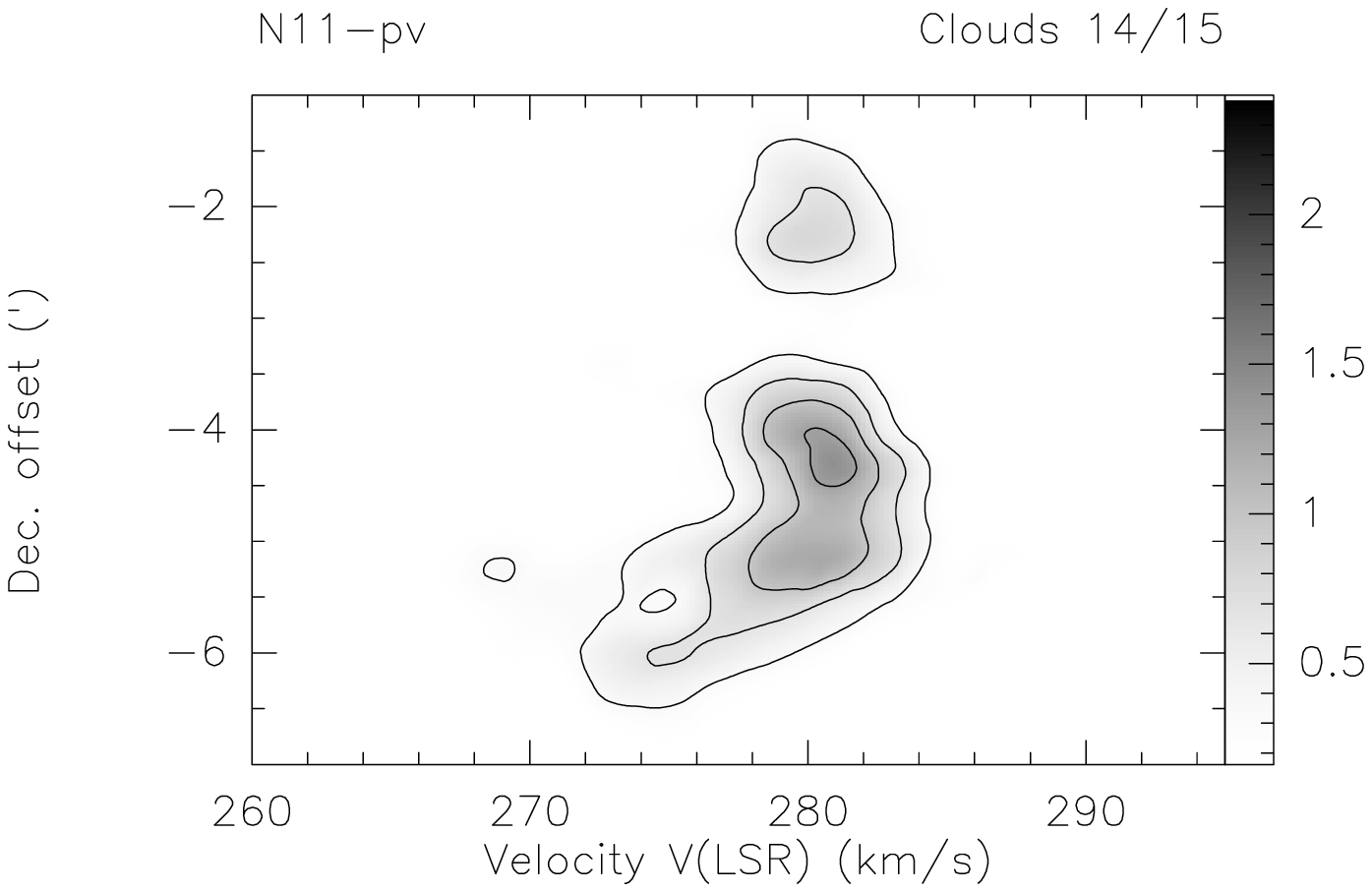}}}
\end{minipage}
\end{minipage}
\caption{$J$=1-0 $\co$ position-velocity maps for clouds 1 through 15
associated with the `ring' in N~11. All maps are at constant right ascension,
except those of clouds 8, 9, 10 and 11 which are at constant declination. 
Contours are multiples of $\int T_{\rm mb}{\rm d}V = 0.46 \kkms$. 
Grey scales are labeled in $\int T_{\rm A}^{*}{\rm d}V$.
}
\label{n11pv}
\end{figure*}

\begin{figure*}[]
\addtocounter{figure}{-1}
\unitlength1cm
\begin{minipage}[b]{17.5cm}
\resizebox{17.5cm}{!}{\rotatebox{0}{\includegraphics*{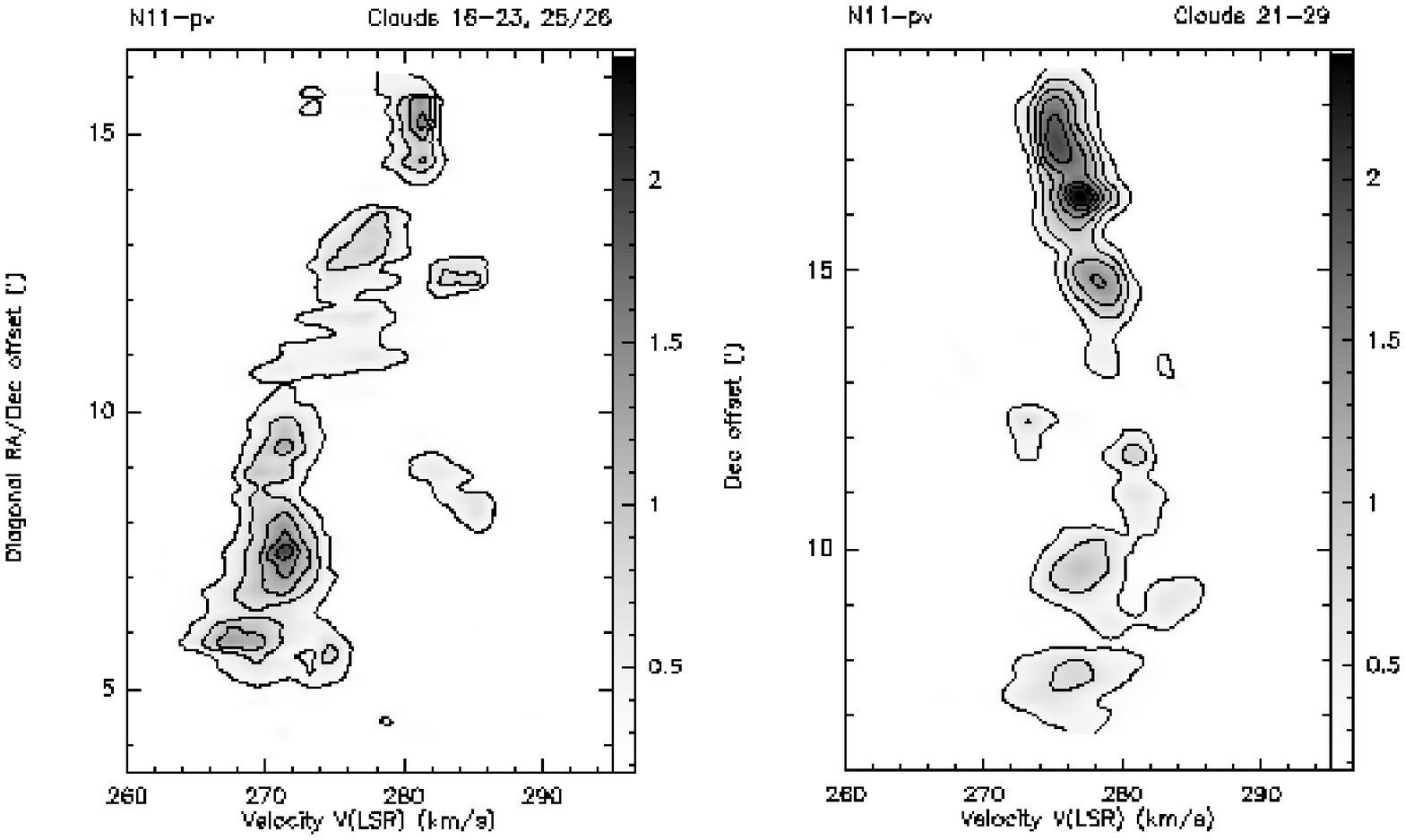}}}
\end{minipage}
\caption{continued; $J$=1-0 $\co$ position-velocity maps for clouds 16
through 29 associated with the northeastern `chain' in N~11. The map
for clouds 16-23 is along a diagonal originating at the map (0,0)
position and extending to the northeast in position angle 45$^{\circ}$
(counterclockwise). The map for clouds 21-29 is at constant right ascension.
Contours are at multiples of $\int T_{\rm mb}{\rm d}V = 0.46 \kkms$.
Grey scales are labeled in $\int T_{\rm A}^{*}{\rm d}V$.}
\end{figure*}

\begin{table}
\caption[]{Line ratios of N~11 CO clouds}
\begin{center}
\begin{tabular}{| r | lrl |}
\hline
\noalign{\smallskip}
No. &  $\textstyle{J=2-1\over J=1-0}$ & $\textstyle{\co\over \13co}$ & $\textstyle{\co\over \13co}$ \\
    &  $\co$ & $J$=1--0 & $J$=2--1 \\
    &        &          & \\
\noalign{\smallskip}	
\hline			
\noalign{\smallskip}	
  4 & 0.7$\pm$0.2       &  9.4$\pm$1.4 & \\
  5 &                   & 14.2$\pm$2.1 & \\
  7 &                   &  5.1$\pm$1.1 & \\
  8 & 1.2$\pm$0.3       &  7.4$\pm$1.1 & \\
 10 & 1.2$\pm$0.3$^{a}$ &  9.8$\pm$1.2 & 4.8$\pm$0.9 \\
 11 & 0.9$\pm$0.2$^{a}$ &  5.3$\pm$0.8 & \\
 12 &                   &  7.9$\pm$1.1 & \\
 13 & 0.7$\pm$0.3$^{a}$ &  8.8$\pm$2.8 & \\
 14 & 1.3$\pm$0.4$^{a}$ &  6.0$\pm$0.8 & 4.7$\pm$0.9\\
 15 & 1.2$\pm$0.2$^{a}$ &  7.6$\pm$1.1 & \\
 18 & 1.1$\pm$0.2       & 22.8$\pm$3.7 & \\
 27 &                   & 15.7$\pm$2.0 & \\
 28 &                   & 10.1$\pm$1.2 & \\
 29 &                   &  8.4$\pm$1.1 & \\
\noalign{\smallskip}
\hline
\end{tabular}
\end{center}
Notes: $^{a}$ average over cloud
\end{table}

\begin{table}
\caption[]{Properties of CO clouds in N~11}
\begin{center}
\begin{tabular}{| r | rrrr |}
\hline
\noalign{\smallskip}
No & Luminosity & Mean        & Virial                  & $X$ \\
   & $L_{\rm CO}$  & Radius      & Mass                 & 10$^{20}$ cm$^{2}$ \\
   & $\kkms$ pc$^{2}$ &$R$ pc& 10$^{4}$ M$_{\odot}$ & $(\kkms)^{-1}$ \\
\noalign{\smallskip}	
\hline			
\noalign{\smallskip}	
  1 &  2705 & 10.6       &   1.6 &   3.8 \\   
  2 &  2650 &  7.3       &   1.1 &   2.6 \\   
  3 &  2725 & 11.3       &   2.0 &   4.6 \\   
  4 & 10470 & 11.2       &   7.6 &   4.6 \\   
  5 &  3955 &  8.9       &   2.0 &   3.2 \\   
  6 &  5790 & 19.9       &   2.6 &   2.9 \\   
  7 &  3400 & $<$5       &$<$0.4 &$<$0.7 \\
  8 &  1335 & $<$5       &$<$0.7 &$<$3.4 \\
  9 &  1485 &  7.0       &   1.0 &   4.2 \\   
 10 &  5720 &  7.4       &   5.4 &   5.9 \\   
 11 &  2995 &  7.7       &   3.3 &   6.9 \\   
 12 &  2560 &  7.4       &   3.0 &   7.4 \\   
 13 &  2245 & $<$5       &$<$1.0 &$<$2.9 \\
 14 &  1565 & $<$5       &$<$1.7 &$<$6.8 \\
 15 &  9290 & 10.7       &   3.2 &   2.2 \\   
 16 &  2890 &  7.7       &   5.4 &  11.8 \\   
 17 &  1400 &  8.6       &   0.5 &   2.4 \\   
 18 &  4385 &  8.5       &   2.7 &   3.9 \\   
 19 &  2725 &  9.0       &   2.6 &   5.9 \\   
 20 &  1190 & $<$5       &$<$7.2 &$<$38 \\
 21 &  1885 &  6.9       &   3.3 &  11.2 \\   
 22 &   850 &  6.0       &   0.5 &   3.7 \\   
 23 &  3505 & ---        &   --- &  ---  \\   
 24 &  1660 &  8.3       &   1.7 &   6.3 \\   
 25 &   510 & $<$5       &$<$1.0 &$<$12  \\
 26 &  1925 & 10.4       &   0.9 &   2.8 \\   
 27 &  1970 &  6.1       &   1.6 &   5.0 \\   
 28 &  3915 &  7.4       &   2.0 &   3.2 \\   
 29 &  2615 & $<$5       &$<$1.1 &$<$2.6 \\
\noalign{\smallskip}
\hline
\end{tabular}
\end{center}
\end{table}

N~11 is prominent not only at optical wavelengths, but also in the infrared
and radio continua (Schwering $\&$ Israel 1990; Haynes et al. 1991) and in 
CO line emission (Cohen et al. 1988). It has a complex structure (see 
Fig.~\ref{n11overall}). The southern part of N11 appears to be a 
filamentary shell of diameter 200 pc enclosing the OB association LH~9 
(Lucke $\&$ Hodge 1970) also known as NGC 1760. In the center of this
shell, we find the relatively inconspicuous HII-region N~11F. At the 
northern rim of the shell, another OB association, LH~10 (a.k.a. 
NGC 1763, IC 2115, IC 2116) is associated with the very bright HII region 
N~11B and the bright, compact object N~11A (Heydari-Malayeri $\&$ Testor 
1983). The eastern rim of the shell is likewise marked by the OB 
association LH~13 (NGC 1769) exciting the bright HII regions N~11C and
N~11D. Finally, OB association LH~14 (NGC 1773), coincident with HII 
region N~11E, marks the point where the northeastern loop SGS-1 meets 
the filamentary shell around LH~9. The HII regions N~11B, N~11CD, N~11E
and N~11F are all identified with thermal radio sources in the catalog
published by Filipovic et al. (1996). The far-infrared emission from
warm dust does not show the same spatial distribution as the radio 
continuum and H$\alpha$ line emission from ionized hydrogen gas 
(see Fig.~3 in Xu et al. 1992). The latter fills the whole shell
region, whereas the former is clearly enhanced at the shell edges.
The radio HII regions have typical r.m.s. electron-densities of
15 $\cc$, masses of $10^{4}-10^{5}$ M$_{\odot}$, emission measures
of $10^{4}$ pc cm$^{-6}$ and appear to be well-evolved (Israel
1980). The OB associations powering the complex are all rich associations.
For instance, LH~9 contains 28 O stars, and LH~10 contains 24 O stars
(Parker et al. 1992). Likewise, LH~13 contains some 20 O stars, and
LH~14 about a dozen (Heydari-Malayeri et al. 1987). It is possble
that star formation in the N~11 complex is at least partly triggered
by the expanding shell surrounding LH~9 (Rosado et al. 1996).

The low-resolution (12$'$) CO observations by Cohen et al. (1988) showed 
that the N~11 group of HII regions is associated with an extended molecular 
cloud complex. It is the third brightest CO source in their survey, after 
the very extended 30 Doradus complex, and the more modest N~44 complex.
Cohen et al. estimated for the N~11 molecular complex a mass of about 
$M(\h2) = 5 \times 10^{6}$ M$_{\odot}$, although the comparison of these
data with IRAS results by Israel (1997) suggests about half this value. 
A higher-resolution (2.6$'$) CO survey, carried out by Mizuno et al. 
(2001) had insufficient sensitivity to reproduce the actual CO structure;
the low (virial) mass estimate given appears to be rather uncertain.
Using the same instrument, Yamaguchi et al. (2001) conducted a more 
sensitive survey in which extended emission from a CO cloud complex
is seen to follow the outline of the ionized gas making up the HII  
region complex.

Because of its prominence and its interesting optical structure, we have
mapped N~11 and its surroundings in the $J$=1--0 $\co$ transition within
the framework of the ESO-SEST Key Programme. Preliminary results have
been presented by Israel $\&$ de Graauw (1991) and Caldwell $\&$ Kutner
(1996). We have also mapped parts of the complex in the $J$=2--1 $\co$ 
transition, and in the corresponding transitions of $\13co$.

\section{Observations}

\begin{figure*}[]
\unitlength1cm
\begin{minipage}[b]{17.75cm}
\resizebox{17.5cm}{!}{\rotatebox{0}{\includegraphics*{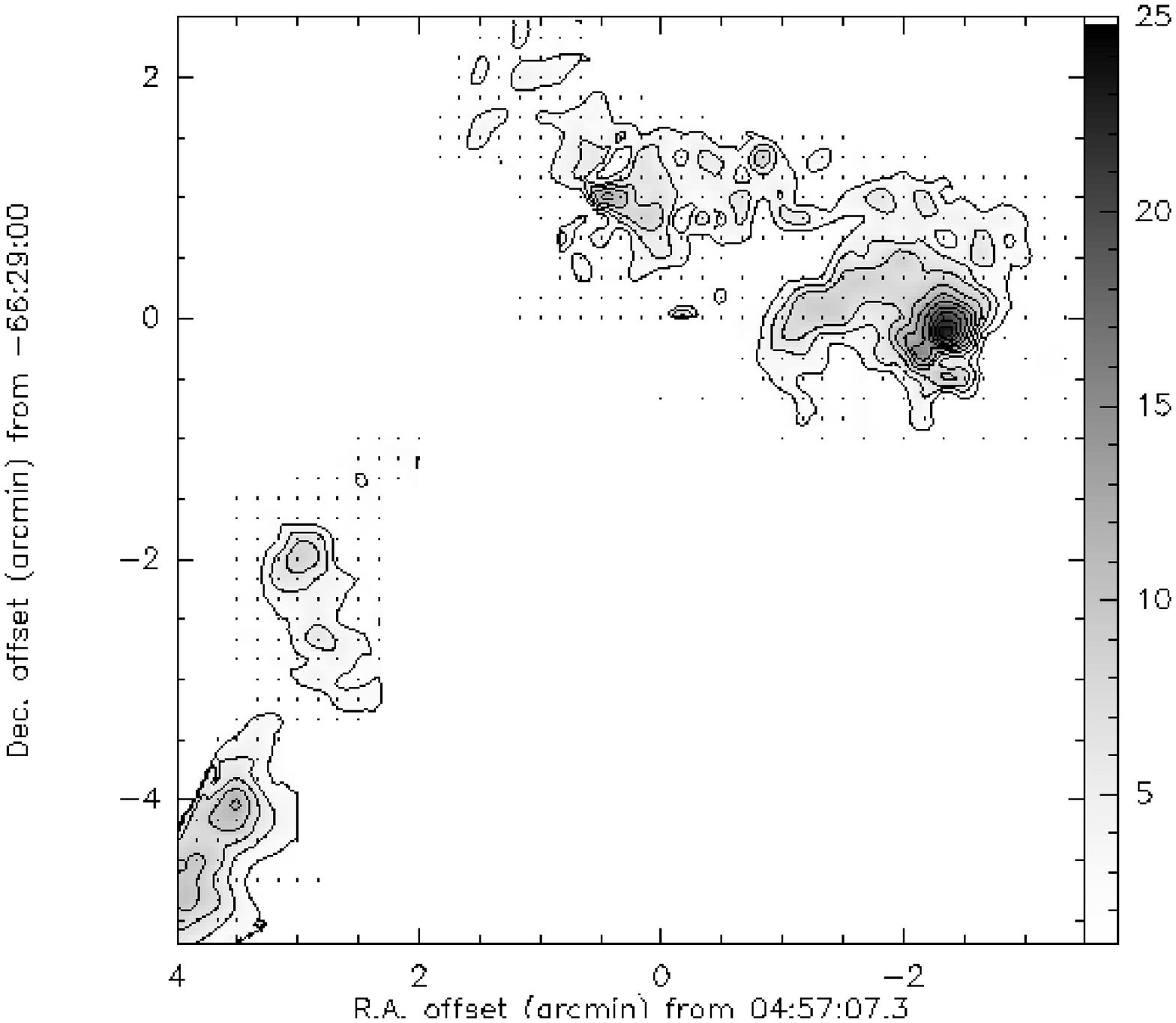}}}
\end {minipage}
\caption{Velocity-integrated $J$=2-1 $\co$ map of the northeastern segment 
of the ring, containing clouds 10, 11, 13, 14 and 15. Contours are at 
multiples of $\int T_{\rm mb}{\rm d}V = 4 \kkms$.
Grey scales are labeled in $\int T_{\rm A}^{*}{\rm d}V$.}
\label{230map}
\end{figure*}

The $\co$ (1--0) observations were mostly made in a single observing run
in December 1988 and January 1989 using the SEST 15 m located on 
La Silla (Chile)\footnote{The Swedish-ESO Submillimetre Telescope 
(SEST) is operated jointly by the European Southern Observatory (ESO) 
and the Swedish Science Research Council (NFR).}. Smaller data sets
obtained in April 1988 and in October 1989 were also used.
The $\co$ (2-1) measurements were made during four runs in 1989, 1992,
1993 and 1994. Although some $\13co$ $J$=1--0 observations had already
been made in 1988, most were obtained during the 1993 and 1994 runs;
the relatively few $\13co$ $J$=2--1 observation were all made in the 
1992 run. All $J$=1--0 observations were made with a Schottky 
receiver, yielding typical overall system temperatures $T_{\rm sys}$ = 
600 -- 750 K. The $J$=2--1 observations were made with an SIS mixer, 
yielding typical overall system temperatures $T_{\rm sys}$ = 450 -- 
750 K depending on weather conditions. On average, we obtained 1$\sigma$
noise figures in a 1 km s$^{-1}$ band of 0.04, 0.10, 0.08 and 0.12 K 
at 110, 115, 220 and 230 GHz respectively.

In both frequency ranges, we used the high resolution acousto-optical 
spectrometers with a channel separation of 43 kHz. The $J$=1--0
observations were made in frequency-switching mode, initially (1988) 
with a throw of 25 MHz, but subsequently with a throw of 15 MHz. 
The $J$=2--1 measurements were made in double beam-switching mode, 
with a throw of 12$'$ to positions verified from the $J$=1--0
$\co$ map to be free of emission. Antenna pointing was checked 
frequently on the SiO maser star R Dor, about 20$^{\circ}$ from the
LMC; r.m.s. pointing was about $3''-4''$. The N~11 area was first
roughly sampled in the $J$=1--0 $\co$ transition on a grid of $80''$ 
(double-beam) spacings, using IRAS infrared maps (Schwering $\&$ Israel 
1990) as a guide. Where emission was detected, we refined the grids to a 
half-beam sampling of $20''$. Some of the clouds thus mapped in $J$=1--0 
$\co$ were observed in $J$=1--0 $\13co$ on the same grid, and with
$10''$ grid-spacing in the $J$=2--1 transitions.

\begin{figure*}[]
\unitlength1cm
\begin{minipage}[b]{17.5cm}
\resizebox{17.0cm}{!}{\rotatebox{0}{\includegraphics*{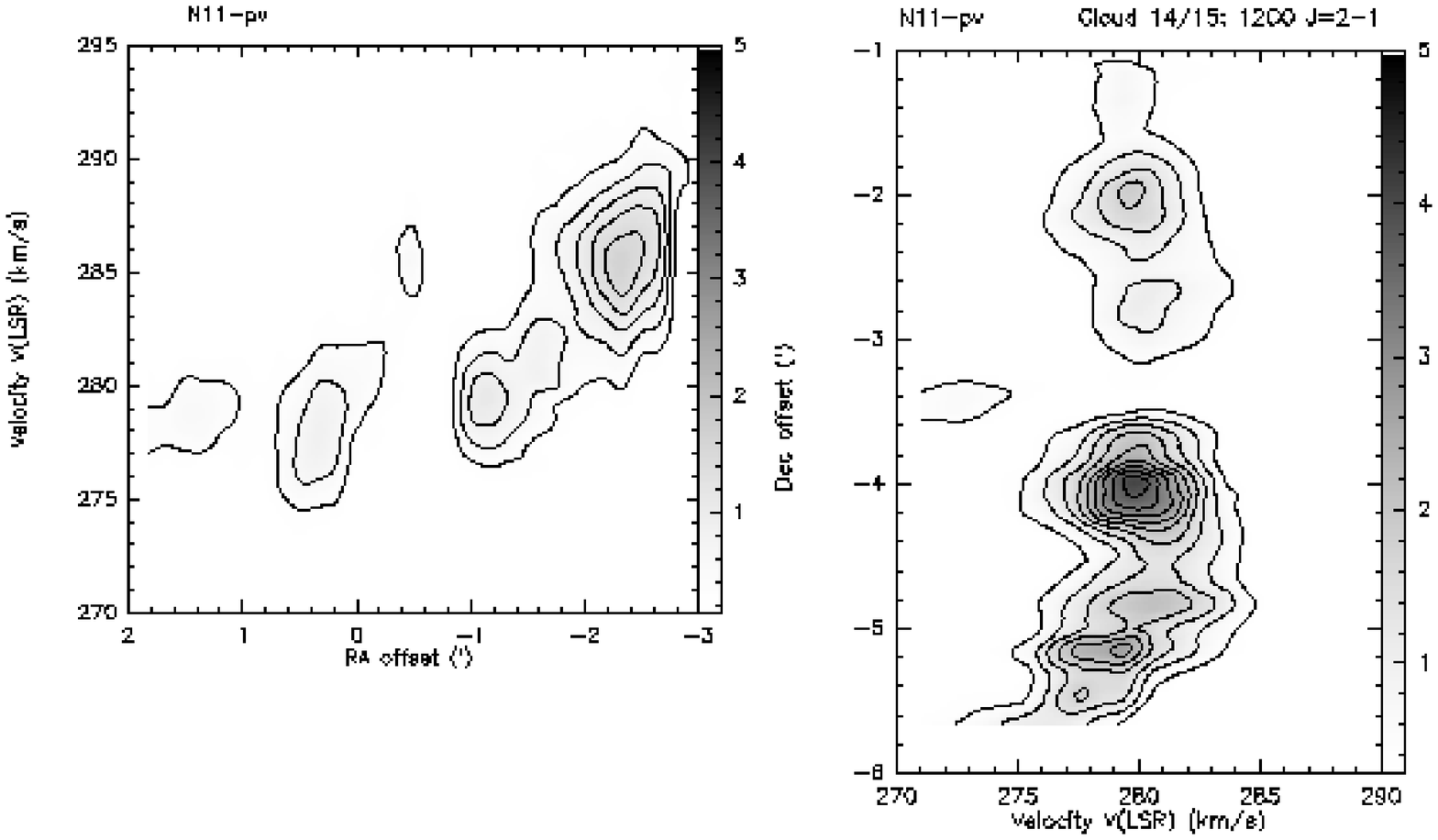}}}
\end{minipage}
\caption{Position-velocity maps through clouds depicted in 
Fig.~\ref{230map}. Left: right-ascension cut at constant
declination through clouds 10 and 11, with contours at multiples of 
$\int T_{\rm mb}{\rm d}V = 0.5 \kkms$. Emission was averaged over
1 arcmin in declination. Velocity resolution $ 1 \kms$.
Right: declination cut at constant right ascension through clouds 14 and 
15, with contours at multiples of $\int T_{\rm mb}{\rm d}V = 0.75 \kkms$. 
Emission was averaged over 1 arcmin in right ascension. Velocity 
resolution $ 2 \kms$. In both panels, grey scales are labeled in $\int 
T_{\rm A}^{*}{\rm d}V$.}
\label{230pv}
\end{figure*}

Unfortunately, frequency-switched spectra suffer from significant baseline
curvature. For N11 we corrected baselines by fitting polynomials to the 
baselines, excluding the range of velocities covered by emission and the 
ranges influenced by negative reference features. The emission velocity
range was determined by summing all observations. Inspection by eye suggested
that this method worked well. It has the advantage that, in principle, it
does not select against weak extended emission, as long as this covers the
same velocity range as the brighter clouds.

The FWHM beams of the SEST are $45''$ and $23''$ respectively at frequencies
of 115 GHz and 230 GHz. Nominal main-beam efficiencies $\eta_{\rm mb}$ at 
these frequencies were 0.72 and 0.57 respectively. For a somewhat more
detailed discussion of the various efficiencies involved, we refer to
Johansson et al. (1998; Paper VII). 

\section{Results and analysis}

\subsection{Catalogue of CO clouds}

An overview of the $J$=1--0 $\co$ mapping results is shown in 
Fig.~\ref{n11overall}, directly comparable to the sketch of optical emission. 
More detailed maps of the southwestern and northeastern parts of the N~11 
complex are shown in Figs.~\ref{n11sw} and ~\ref{n11ne} respectively. 
Kinematical information is represented by channel maps in Fig.~\ref{n11chans} 
and position-velocity maps along selected cuts in Fig.~\ref{n11pv}. The 
distribution of CO emission in the N~11 complex is remarkable. Using both 
position and kinematical information, at least 29 well-defined individual 
clouds can be identified. The actual number of clouds is higher than this. 
For instance,  the velocity widths of clouds 4, 16, 20, 23, and perhaps 
cloud 10 as well, suggest that clouds with different velocities, but in 
the same line of sight, are blended together. Moreover, in the sparsely 
sampled parts of the map, clouds with relatively weak emission may have 
escaped our attention. For instance, inspection of individual profiles 
reveals that weak, but significant emission (typically $T_{\rm mb} \approx 
0.3$ K, $\int{T_{\rm mb}{\rm d}V \approx 2 \kkms}$) is present at some
positions in the map. This is the case just outside the southwestern edge 
of the `ring' at positions (-10.7, -14.7) and (-13.3, -12), velocity 
$V_{\rm LSR} = 274 \kms$, outside the southeastern edge at position 
(4, -14.7), velocity $V_{\rm LSR} = 277 \kms$ and inside the `ring' at 
(0, -6) with $V_{\rm LSR} \approx 276 \kms$. In the `empty' southeastern 
part of the ring, very weak emission is likewise found at velocities 
between 274 and 284 $\kms$, whereas stronger emission ($T_{\rm mb} 
\approx 0.4$ K) occurs in the gap between clouds 5 and 10, at velocities 
of 267 and 280 $\kms$. Finally, extended weak emission appears to be 
present around (-7, -9) with $V_{\rm LSR} = 272 \kms$. 

All well-defined clouds and their observational properties, are listed
in Table 1 which also identifies the corresponding IRAS infrared source 
and radio continuum sources from the catalogues by Schwering $\&$ Israel 
(1990) and Filipovic et al. (1996). For each cloud, we give the central 
position and the parameters of the peak antenna temperature $J$=1--0 and
$J$=2--1 $\co$ profiles. Clouds can readily be identified by
referring the position in Table 1 to Fig.~\ref{n11overall}.

\subsection{Lack of diffuse emission}

The appearance of N~11 is rather different from that presented by 
cloud complexes in quiescent, non-star-forming regions of the LMC, 
such as the cloud complexes discussed in Paper VI (Kutner et al. 
1997): compare in particular our Fig.~4 with their Fig.~4. In the
latter, long chains of individual bright clouds are connected
by continuous, relatively bright intercloud emission. The N~11 map
is dominated by discrete clouds. More extended, diffuse intercloud 
emission is almost wholly absent, as already noted by Caldwell $\&$
Kutner (1996). By summing emission from many 
`empty' positions, we have found that there is no diffuse emission 
above  $T_{\rm mb} \approx 0.07$ K anywhere in the southwestern part of 
the N~11 complex (Fig.~\ref{n11sw}), so that the clouds in the `ring' region
thus have a very high contrast with their surroundings. Some amount 
of diffuse emission is present in the chain of clouds extending to the 
northeast (cf. Fig.~\ref{n11ne}).  We may quantify the lack of diffuse 
emission by comparing the sum of the individual cloud CO luminosities in 
Table~1 ($\Sigma {L_{\rm CO} = 9.0 \times 10^{4} \kkms {\rm pc}^{2}}$) to 
the independently determined integral CO luminosity from  
the {\it whole} N~11 map ($\int{L_{\rm CO} = 11.0 \times 10^{4} \kkms 
{\rm pc}^{2}}$). It thus appears that, overall, the {\it identified 
discrete CO clouds} alone provide already 82$\%$ of the total CO 
emission. As may be surmised from the above, the fractions are 
different for the southwestern ring region and the northeastern chain. 
For these map areas, we find values of 93$\%$ and 75$\%$ respectively.
This means that, in an absolute sense, the northeastern chain contains
twice as much diffuse CO as the ring region.

\subsection{Individual cloud properties}

Although 22 out of 29 of the clouds listed in Table~1 are resolved, 
virtually all of them have dimensions no more than a few times the size 
of the $J$=1--0 $\co$ observing beam. The maps in Figs.~\ref{n11overall} 
through ~\ref{n11pv} therefore do not provide much information on the 
actual structure of individual clouds. In order to determine cloud CO 
luminosities and mean radii, we made for each cloud a small map (not shown) 
over the relevant range of positions and velocity. Cloud CO luminosities 
were determined by integrating these maps. We verified that the results
were not significantly affected by the precise size and velocity limits 
of the maps. Characteristic cloud dimensions were determined by dividing 
the map integral by the map peak and taking the square root. The radii 
thus obtained were then corrected for finite beamwidth, the beam FWHM 
diameter of 43$''$ corresponding to a linear diameter of 11.2 pc.   

The twice higher angular resolution of the $J$=2--1 $\co$ maps shown in 
Fig.~\ref{230map} and ~\ref{230pv} does provide some structural information
at least for clouds 10, 11, 13, 14 and 15 in the relatively bright 
northeastern segment of the ring. In all cases, the cloud structure 
thus revealed is one of mostly low-brightness CO emission with the
confinements of the $J$=1--0 source extent in which a few essentially 
unresolved components are embedded. The overall extent of these compact 
components is thus significantly less than 5 pc.

\begin{figure*}
\unitlength1cm
\resizebox{6.2cm}{!}{\rotatebox{0}{\includegraphics*{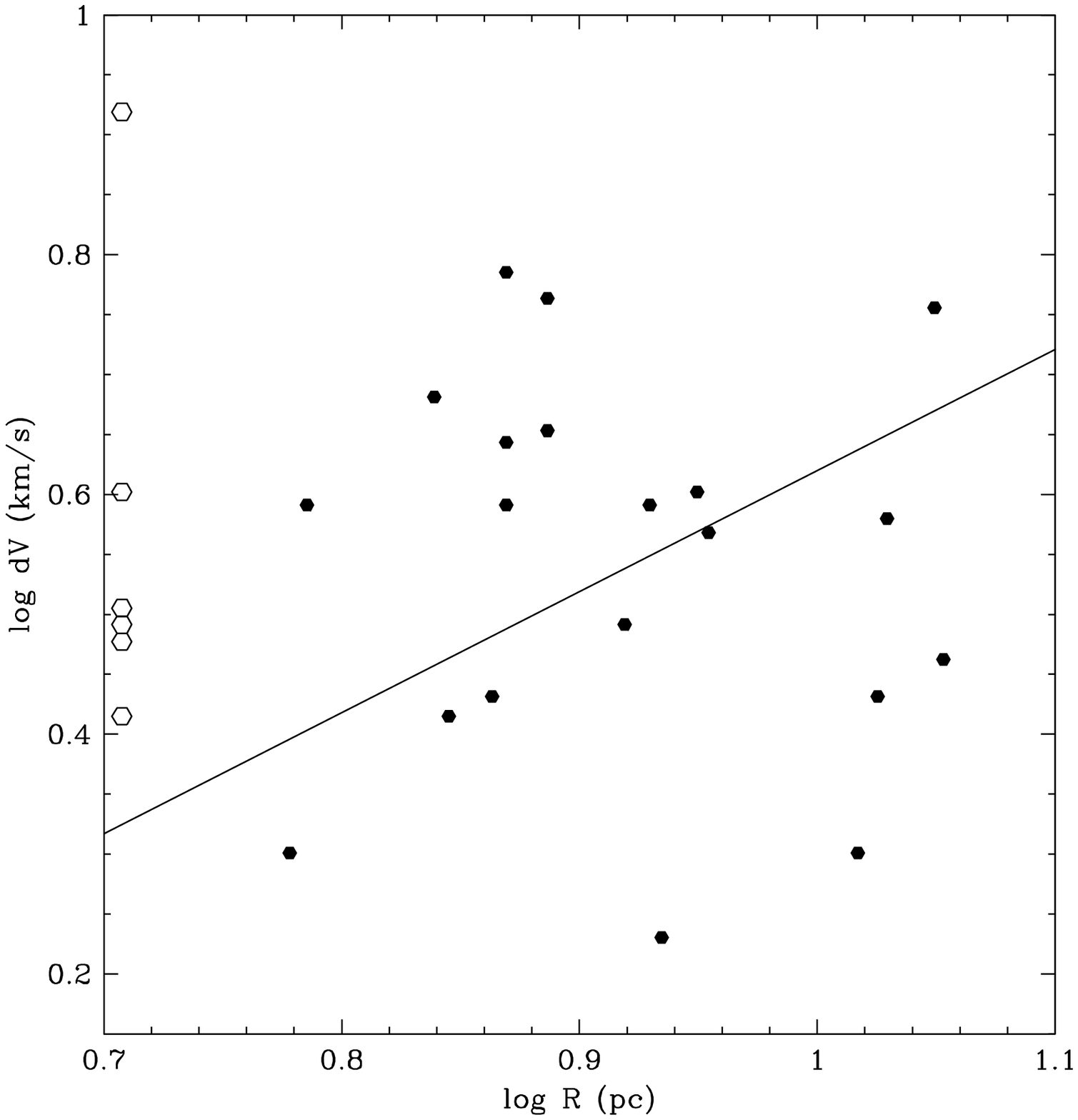}}}
\resizebox{6.2cm}{!}{\rotatebox{0}{\includegraphics*{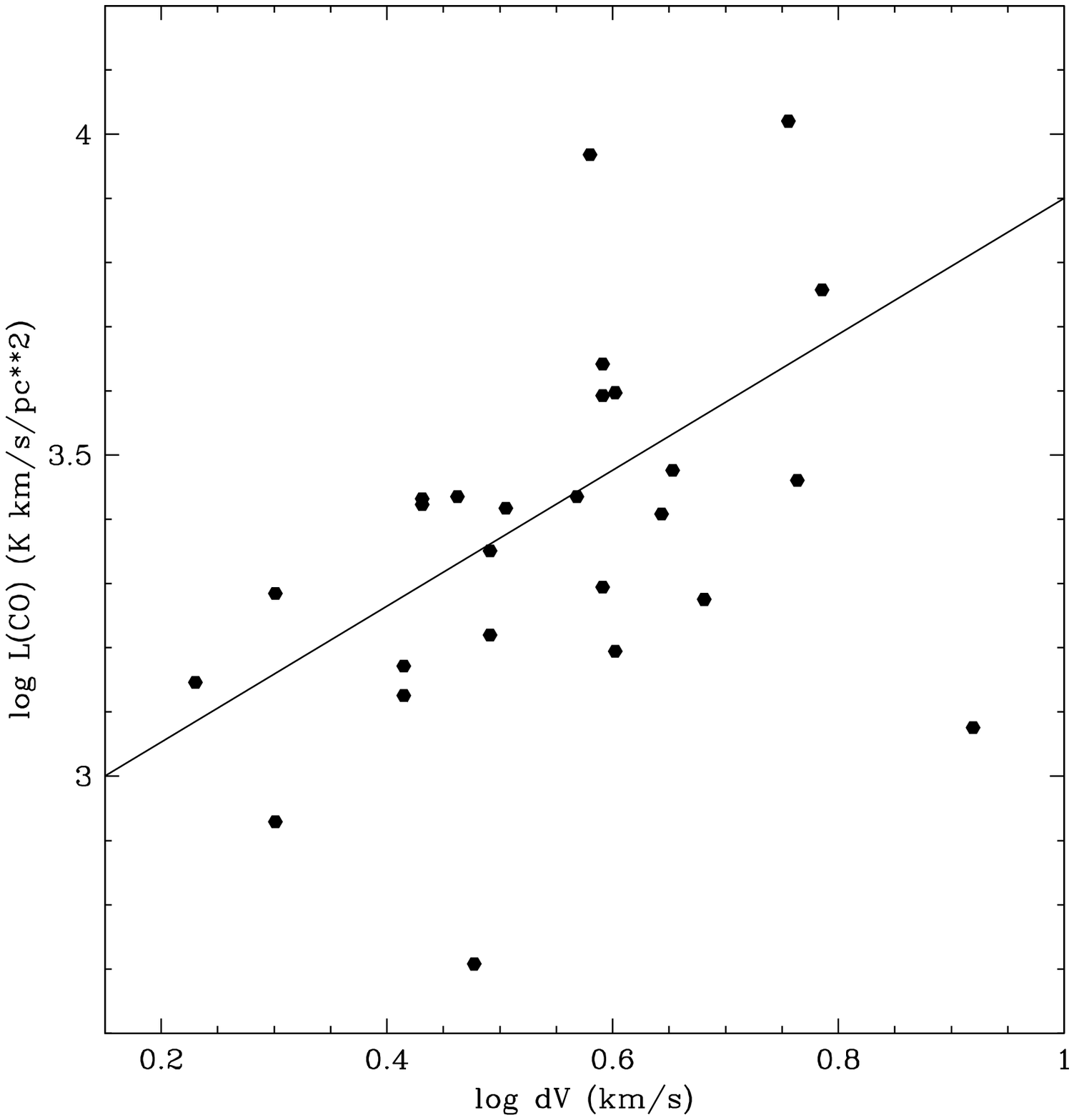}}}
\resizebox{6.2cm}{!}{\rotatebox{0}{\includegraphics*{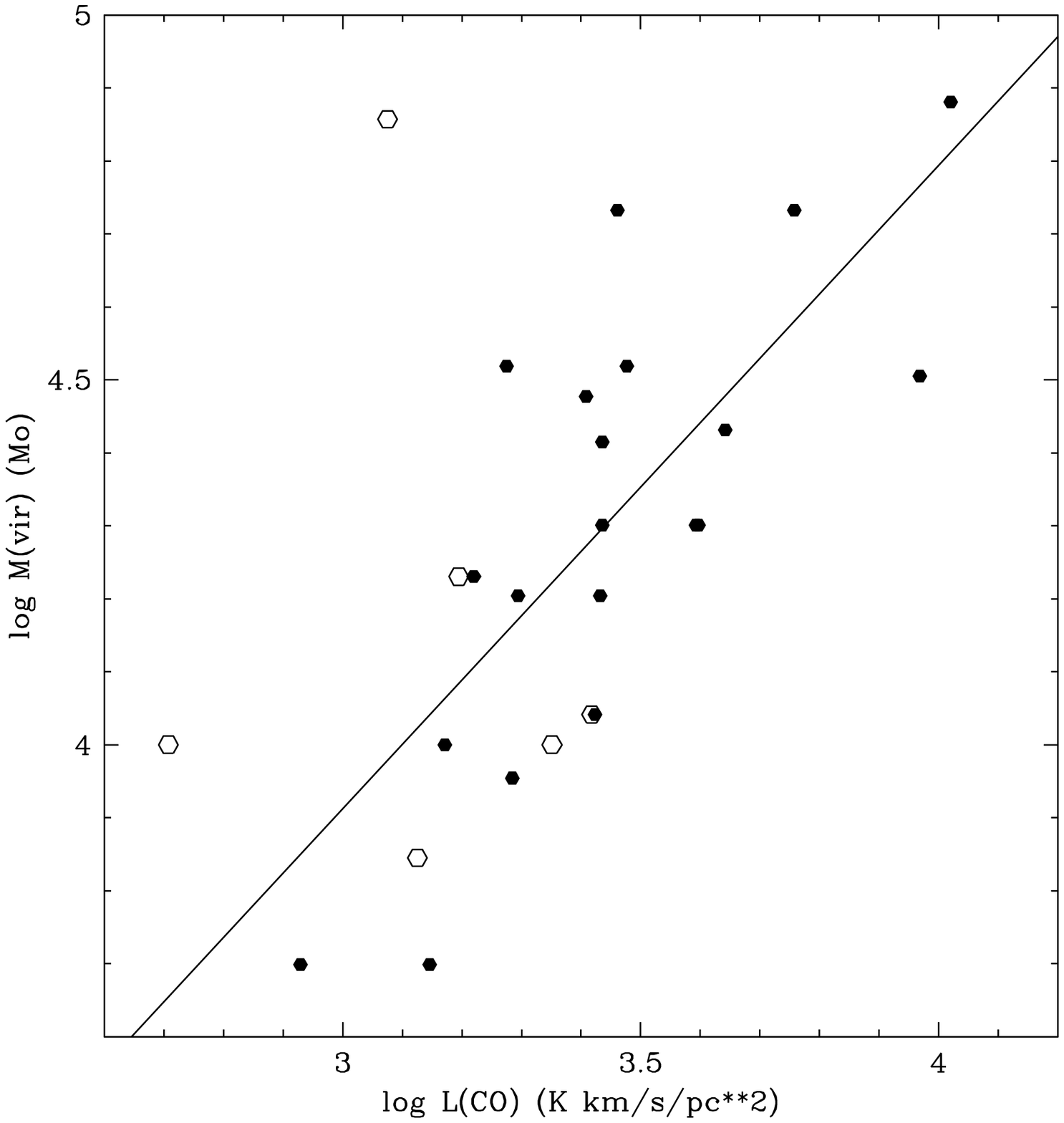}}}
\caption[]{The various parameters of the N~11 cloud ensemble do not
appear to define very clear relationships. Left: Cloud velocity 
width d$V$ as a function of deconvolved radius $R$. Center: CO 
luminosity $L_{\rm CO}$ as a function of deconvolved radius 
$R$. Right: Virial mass $M_{\rm vir}$ as a function of CO luminosity
 $L_{\rm CO}$. Filled hexagons represent actual values, open hexagons 
represent upper limits to cloud size (left panel) or to cloud virial
mass (right panel).
}
\label{n11slum}
\end{figure*} 

Although it is by no means certain that the clouds identified by us 
are indeed virialized, we have used the data given in Tables~1 and 3 
to calculate virial masses following:\\

\noindent
$M_{\rm vir}/{\rm M_{\odot}} = {\rm k}\, R/{\rm pc}\, (\Delta V/{\kms})^{2}$\\

\noindent
where k = 210 for homogeneous spherical clouds and k = 190 for clouds
with density distributions $\propto r^{-1}$  (MacLaren et al. 1988).
In our calculations, we have assumed the former case, although the 
actual uncertainties are in any case much larger than the difference 
between the two values of k. The results are included in Table~3.

In previous papers we have found that cloud size, linewidth, 
luminosity and virial mass appear to be related quantities, 
although the precise form of the relations is different for SMC
(Rubio et al. 1993) and LMC (Garay et al. 2002). Using data from
Tables~1 and 3, we have investigated these relations also for the 
clouds in the N~11 complex. Fig.~\ref{n11slum} we present the 
results, including the formal least-squares fits :\\

\noindent
log d$V$ = 1.01 log $R$ -- 0.39, \hfill \\

\noindent
log $L_{\rm CO}$ = 0.94 log d$V$  + 2.96\\

\noindent
log $M_{\rm vir}$ = 0.88 log $L_{\rm CO}$ + 1.27 \\

\noindent

Compared to the clouds studied before, those in N~11 complex 
are characterized by a relatively limited span in all 
parameters. The velocity width and linear size in particular 
cover only a very limited range. In addition, the diagrams in 
Fig.~\ref{n11slum} exhibit a relatively large and mostly 
intrinsic scatter of data points. Not surprisingly, therefore, 
are the very small determination coefficients $r^{2} \approx 
0.1$ of the d$V-R$ (left panel) and $L_{\rm CO}-$d$V$ (center 
panel) relations. Thus, unlike the clouds studied before in 
both the LMC and the SMC, the N~11 clouds do not appear to 
exhibit a significant correlation between these quantities.

However, the relation between cloud virial mass and and CO 
luminosity (right panel) seems better defined, with a 
determination coefficient $r^{2} \approx 0.6$. We note that 
the regression fit, shown in Fig.~\ref{n11slum}, has
a slope very similar to the one found by Garay et al. (2002)
for clouds in 30 Doradus and the surrounding LMC environment 
(and rather different from the one obtained for SMC clouds by 
Rubio et al. 1993). At the same time, for identical CO 
luminosities cloud virial masses are systematically $\it lower$ in 
the N~11 complex than in the Doradus clouds by a factor of 2.5. 
Leaving aside any speculation as to the origin of these
differences, we feel confident to conclude that the N~11 
cloud properties differ significantly from those studied at 
other locations either within the LMC or the SMC.

\subsection{CO cloud physical condition}

We have determined $J$=1--0 $\co/\13co$ line ratios for half the 
clouds listed in Table~1. These ratios were usually measured near 
but not precisely at the integrated $\co$ peaks. Moreover, in
most cases the ratio was measured at various positions. The 
intensity-weighted means of these measurements and their errors
are listed in Table~2. Individual values for this ratio (which
we will call the isotopical ratio) range from 5 to 23, with a mean 
of about 10 (Table~2). Two similar determinations in the $J$=2--1 
transition yield a value of about 5. 

For five clouds we could integrate the $\co$ emission in the 
$J$=1--0 and $J$=2--1 transitions over identical areas, thus
obtaining the $J$=2--1/$J$=1--0 line ratio (i.e the transitional 
ratio) listed in Table 2 as `average over cloud'. for another
three clouds we mapped small crosses in the $J$=2--1 transition,
allowing us to extrapolate to the larger $J$=1--0 beam size.
The transitional ratios thus derived are typically 1.2. However,
the bright clouds 4 and 13 are exceptional in having the much 
lower value of only 0.7. 

We have run radiative transfer models  
as described by Jansen (1995) and Jansen et al. (1994) in an attempt
to reproduce the observed ratios as a function of input parameters
such as molecular gas kinetic temperature $T_{\rm k}$, molecular
hydrogen gas density $n_{\h2}$ and CO column density per unit
velocity $N(CO)$/d$V$. Although the models assume a homogeneous, 
plane-parallel geometry, this is an acceptable approximation .

Because three model parameters are required,
the solutions are poorly constrained, except in the cases of clouds
10 and 14, where three line intensity ratios are available for
fitting. We find that cloud 10 is best fit by a moderately dense
($n_{\h2} = 3000 \cc$), hot ($T_{\rm k}$ = 150 K) molecular cloud
with a CO column density $N(CO)$/d$V$ = $6 \times 10^{17} \cm2 
(\kms)^{-1}$ and a surface filling factor of 0.04 (see also, for
instance, Rubio et al. 2000). In contrast,
the overall beam surface filling factor is about 0.25. 
Although the model transitional ratio is 0.85 instead of 
the observed value 1.2$\pm$0.2, the model isotopical ratios are
practically identical to those observed. Only one other model solution 
comes close to the observed value. It provides a poorer fit and
requires very high densities ($n_{\h2} = 100 000 \cc$) and very low
column densities $N(CO)$/d$V$ = $0.3 \times 10^{17} \cm2 (\kms)^{-1}$ at 
low temperatures ($T_{\rm k}$ = 10 K). As Cloud 10 is very closely 
associated with the rich and young OB association LH~10 and the
bright HII region N~11B, we consider the parameters of the first 
solution to be more likely correct. 

However, it is unlikely that all of cloud 10 is both hot and dense.
Whether or not the cloud is virialized, we expect its mass not to
be very different from the value $M = 5 \times 10^{4}$ M$_{\rm \odot}$
given in Table~3. To heat all of that mass to a temperature of 150 K
is beyond the capacity of the OB association, even if a large fraction
of it is embedded and not yet properly identified. Rather, we suspect
that cloud 10 is characterized by a range of temperatures and
densities, with $\co$ emission preferentially dominated by the 
presumably relatively small amounts of hot gas, while the $\13co$ 
intensities are more susceptible to more widespread denser gas. 
Although the present observations do not allow fitting of such a
multi-component model, future observations of higher $\co$ and $\13co$
transitions will make this easily possible.

Clouds 14 and 15 are less closely associated with the OB association
LH~13 and the HII regions N~11C/N~11D. The model solution that
provides the best fit requires again moderate densities  ($n_{\h2} = 
3000 \cc$) and moderate temperatures ($T_{\rm k}$ = 60 K), together
with a slightly lower column density  $N(CO)$/d$V$ = $3 \times 10^{17} 
\cm2 (\kms)^{-1}$ and a surface filling factor of about 0.08. The overall
beam surface filling factor is of the order of 0.3. Other solutions 
found, yielding higher temperatures at lower densities, and vice 
versa, again provide poorer fits. The temperature and mass constraints 
for Clouds 14 and 15 are not as stringent as those for Cloud 10, but
the same comment should also apply to them.

Finally, although the lack of information does not properly
constrain possible solutions, the rather high $J$=1--0 isotopical
ratio of 23 for Cloud 18 does suggest a combination of relatively
high densities and temperatures.

\subsection{Molecular gas mass}

There are various ways in which to estimate the total molecular ($\h2$) 
mass from CO observations. Unfortunately, it is doubtful which of these,
if any, is applicable to N~11. The presence of so many early-type
stars in the immediate vicinity of the molecular material leads one
to suspect that the resulting strong radiation fields have led to
considerable processing of the molecular interstellar medium in N~11.
The observations appear to bear this out: the lack of diffuse CO,
as well as the large and apparently intrinsic scatter in the log 
d$V$--log$R$ and log$L_{\rm CO}$--log d$V$ diagrams (Fig.~\ref{n11slum}),
the various manner in which the detected CO clouds are associated
with FIR dust emission (cf. Table~1) and the elevated temperatures
found above all suggest that in this complex CO has been subject to
different but considerable degrees of photo-processing and 
photo-dissociation. 

The kinematics of the clouds do not suggest regular rotation, or
any other systematic movement, precluding a dynamical mass
determination. For the same reason, it is very difficult to
relate the present results to the overall structures such as 
shells etc that may have resulted from the interaction of the 
many OB stars in the region with the ambient interstellar medium. 
Obviously, the virial theorem cannot be applied to 
the cloud ensemble defining the ring, nor to that forming the 
northeastern ridge of clouds. The way in which the barely 
resolved $J$=1--0 clouds break up in equally barely resolved $J$=2--1 
clouds also casts some doubt on the applicability of the virial 
theorem to the individual clouds, and Fig.~\ref{n11slum} does
not show a very tight relation between virial mass and CO
luminosity. Finally, there is now ample evidence that the `standard'
CO-to-$\h2$ conversion factor $X$ that is often used to derive
molecular hydrogen column densities from CO luminosities is not
valid under the very circumstances pertaining to N~11: strong
radiation fields and low metallicities (Cohen et al. 1988; 
Israel, 1997; see also discussion in Johansson et al. 1998). 

Comparison of the virial masses, corrected for a helium contribution 
of 30$\%$ by mass, with the observed CO luminosity 
supplies the mean CO-to-$\h2$ conversion factor $X$, following: \\

\noindent
$X = 1.0 \times 10^{22}\, R\, (\Delta V^{2})\, L_{\rm CO}^{-1}$ \\

\noindent
The values $X$ thus calculated are also listed in Table 3. We find 
for the {\it discrete CO clouds} a range of $X$ values between $2 
\times 10^{20}$ and $12 \times 10^{20} \cm2 (\kkms)^{-1}$, with a 
mean $X$ = $5\pm0.5 \times 10^{20} \cm2 (\kkms)^{-1}$, i.e. 2.5 
times the `standard' conversion factor in the Solar Neighbourhood. 
Johansson et al. (1998) and Garay et al. (2002) obtained similar 
results for clouds in the 30 Doradus region and Complex 37 respectively. 

These factors can also be compared to those determined independently 
for {\it the whole complex} by Israel (1997, hereafter I97). From a 
comparison of observed far-infrared, HI and CO intensities, i.e. 
explicitly taking all HI in the nebular complex into account, he 
finds $X$(N~11-ring) = $21\pm9 \times 10^{20} \cm2 (\kkms)^{-1}$ and 
$X$(N11-northeast) = $6\pm2 \times 10^{20} \cm2 (\kkms)^{-1}$. As 
discussed by Israel (2000), the conversion factor for whole complexes 
is expected to be higher than that of the individual constituent CO 
clouds. In the latter case, spatial volumes that contain abundant and 
selfshielding $\h2$ but have little or no CO left, are explicitly 
excluded in the virial calculation. the result is thus biased to the 
volumes least affected by photo-processing. Measurements of the whole 
complex avoid such a bias.

The overall conversion factor for the northeast ridge is only 25$\%$ 
higher than the mean for the individual clouds, hardly a significant
difference. Such a value, only a few times higher than the conversion
factor in the Solar Neighbourhood, is characteristic for quiescent 
areas in the moderately low-metallicity LMC (cf. I97), and suggests 
that relatively little processing has taken place in the ridge area. 
The sum of the observed individual cloud masses in the ridge is 
$M_{\rm vir}(ridge) = 3.1 \times 10^{5}$ M$_{\rm \odot}$. Assuming 
no HI to be present {\it in} these clouds but correcting for helium, 
we find from this a total molecular mass for the ridge clouds 
$M_{\h2}(ridge) \geq 2.2 \times 10^{5}$ M$_{\rm \odot}$. Although
this is, strictly speaking, a lower limit because we did not fully 
map the ridge area and additional clouds may have escaped attention,
we note that the more extended maps by Yamaguchi et al. (2001) suggest
that in fact only a very little CO emission occurs outside the area 
mapped by us. 

The data tabulated by I97 imply a neutral hydrogen massa $M_{\rm HI}(ridge) 
\approx 4 \times 10^{5}$ M$_{\rm \odot}$, presumably mostly {\it between} 
the clouds, and  a total molecular hydrogen mass $M_{\h2}(ridge) \approx 
5.5 \times 10^{5}$ M$_{\rm \odot}$. A mass $M_{\h2} \leq 2.5 \times 
10^{5}$ M$_{\rm \odot}$ is unaccounted for by individual clouds,
which should represent molecular material distributed {\it between} the 
CO clouds mapped and not directly observed. We have already found that 
the total CO luminosity observed in the ridge is about a third higher than 
the sum of the individual clouds. Thus, diffuse intercloud CO in the 
ridge will have a luminosity $L_{\rm CO} \geq 1.0 \times 10^{4} \kkms$ 
pc$^{2}$, again a lower limit because of incomplete mapping. 

The situation in the ring is different. The overall conversion factor
(I97) is $5\pm2$ times higher than the mean value for the individual 
clouds. The total CO emission is only a few per cent higher than 
the cloud sum, leaving no more than $L_{\rm CO} \approx 0.5 \times 
10^{4} \kkms$ pc$^{2}$ for the intercloud CO. The $\h2$ mass contained 
in the detected CO clouds is $M_{\h2}(ring) \approx 3.2 \times 10^{5}$ 
M$_{\rm \odot}$, again under the assumption that there is no atomic 
hydrogen contribution to the virial mass. From I97 we find, however, 
total ring-area masses $M_{\h2} = 16.5 \times 10^{5}$ M$_{\rm \odot}$ 
and $M_{\rm HI} = 5.7 \times 10^{5}$ M$_{\rm \odot}$. This result 
therefore predicts the presence {\it molecular} hydrogen not sampled 
by CO ($X_{intercloud} \leq 175 \times 10^{20} \cm2 (\kkms)^{-1}$) 
in amounts of more than twice that of {\it atomic} hydrogen. The ring 
is thus a rather extreme photon-dominated region (PDR), and should 
exhibit characteristic signposts such as strong [CI] and [CII] emission. 

\section{Conclusions}

\begin{enumerate}
\item We have mapped the strong star-forming complex N~11 in
the $J$=1--0 $\co$ line. Additional data were obtained in
the $J$=2--1 $\co$ line, and in the corresponding transitions
of $\13co$. 
\item A total of 29 individual clouds could be identified.
As the N~11 area was not completely mapped and not fully sampled,
the actual number of clouds is probably higher. The clouds
are distributed in a ring or shell surrounding the OB association
LH~9, and in a ridge extending to the northeast which appears
to be associated with the supergiant shell SGS~1.
\item With an 11 pc linear resolution, most of the clouds
are barely resolved. With a twice higher resolution, most
of the emission is seen to come from again barely resolved
structures in these clouds. If the virial theorem is taken 
for guidance, the individual clouds have masses ranging from 
0.5 to 7.5 $\times 10^{4}$ M$_{\rm \odot}$, with a mean of 
$2.5 \times 10^{4}$ M$_{\rm \odot}$.
\item Unlike the apparently quiescent northeastern ridge, 
the ring region appears to be an extreme photon-dominated 
region (PDR). The high overall CO to $\h2$ conversion factor $X$
of this PDR greatly contrasts with the almost 'normal' conversion 
factors of individual dense CO clouds which more effectively
resist the strong UV radiation from the embedded OB associations 
LH~9, LH~10 and LH~13. 
\item There is very little diffuse CO emission between the clouds, 
and indeed in the N~11 complex as a whole. Nevertheless, diffuse
$\h2$ not sampled by CO because of the PDR nature of the complex
should be be present in significant amounts. This is particularly 
true for the ring region. 
\end{enumerate}

\acknowledgements

It is a pleasure to thank the operating personnel of the SEST 
for their support, and Alberto Bolatto for valuable assistance 
in the reduction stage. M.R. wishes to acknowledge support from
FONDECYT through grants No 1990881 and No 7990042.


\begin{thebibliography}{} 
%
\bibitem{} Caldwell D.A., $\&$ Kutner M.L. 1996 \apj 472, 611
\bibitem{} Cohen R.S., Dame T.M., Garay G., et al. 1988 \apjl 331, L95
\bibitem{} Davies R.D., Elliott H.K., $\&$ Meaburn J., 1976 \mnras 81, 89
\bibitem{} Filipovic M.D., Haynes R.F., White G.L., et al., 1996 \auas 
           120, 77
\bibitem{} Garay G., Johansson L.E.B., Nyman L.-$\AA$, et al. 2002 
           \aua in press (Paper VIII)
\bibitem{} Haynes R.F., Klein U., Wayte S.R., et al., 1991 \aua 252, 475
\bibitem{} Henize H., 1956 \apjs 2, 315
\bibitem{} Heydari-Malayeri M. $\&$ Testor G., 1983 \aua 118, 116
\bibitem{} Heydari-Malayeri M., Niemela V.S., $\&$ Testor G., 1987 \aua
           184, 300
\bibitem{} Israel F.P., 1980 \aua 90, 246
\bibitem{} Israel F.P., 1997 \aua 328, 471 (I97)
\bibitem{} Israel F.P., 2000, in: {\it Molecular Hydrogen in Space}, eds.
           F. Combes $\&$ G. Pineau des Forets, Cambridge Unversity
           Press, pp. 293-296
\bibitem{} Israel F.P., $\&$ de Graauw Th., 1991 in `The Magellanic Clouds',
           IAU Symp. 148, Eds. R. Haynes $\&$ D. Milne, Kluwer Acad. Publ.:
	   Dordrecht, p. 45
\bibitem{} Israel F.P. Johansson L.E.B., Lequeux J., et al. 1993 
           \aua 276, 25 (Paper I)
\bibitem{} Jansen D.J., 1995, Ph.D. thesis, Leiden University (NL)
\bibitem{} Jansen D.J., van Dishoeck E.F., $\&$ Black J.H., 1994, 
           \aua 282, 605 
\bibitem{} Johansson L.E.B., Greve A., Booth R.S., et al. 1998 \aua 331, 857
           (Paper VII)
\bibitem{} Kutner M.L., Rubio M., Booth R.S., et al. 1997 \auas 122, 255
           (Paper VI)
\bibitem{} Lucke P.B., $\&$ Hodge, P.W., 1970 \aj 75, 171
\bibitem{} Meaburn J., Laspias V., Solomos N., $\&$ Goudis C., 1989 \aua 
           225, 497
\bibitem{} Meaburn J., 1980 \mnras 192, 365
\bibitem{} Mizuno N., Yamaguchi R., Mizuno A., et al. 2001 \pasj 53, 971
\bibitem{} Parker J.W., Garmany C.D., Massey P., $\&$ Walborn N.R., 1992
           \aj 103, 1205
\bibitem{} Rosado M., Laval A., Le Coarer, E., et al. 1996 \aua 308, 588
\bibitem{} Rubio M., Contursi A., Lequeux J. , et al. 2000 \aua 359, 1139
\bibitem{} Schwering P.B.W., $\&$ Israel F.P., 1990, Atlas and 
           Catalogue of Infrared Sources in the Magellanic Clouds, 
           Kluwer, Dordrecht
\bibitem{} Walker A.R., 1999 in: Post-Hipparcos Cosmic Candles, eds. A. Heck
           $\&$ F. Caputo (Kluwer, Dordrecht), Ap$\&$SpSc Lib. 237, p. 125
\bibitem{} Westerlund B.E., 1990 A$\&$A R 2, 29
\bibitem{} Xu C., Klein U., Meinert D., Wielebinski R., $\&$
           Haynes R.F., 1992 \aua 257, 47
\bibitem{} Yamaguchi R., Mizuno N., Onishi T., Mizuno A., $\&$ Fukui Y.
           2001 \pasj 53, 959sum
 

\end{thebibliography}
\end{document}